\PassOptionsToPackage{unicode}{hyperref}
\PassOptionsToPackage{hyphens}{url}
\PassOptionsToPackage{dvipsnames,svgnames,x11names}{xcolor}
\documentclass[
12pt]{article}
\usepackage{amsmath,amssymb}
\usepackage{setspace}
\usepackage{iftex}
\ifPDFTeX
\usepackage[T1]{fontenc}
\usepackage[utf8]{inputenc}
\usepackage{textcomp} 
\else 
\usepackage{unicode-math}
\defaultfontfeatures{Scale=MatchLowercase}
\defaultfontfeatures[\rmfamily]{Ligatures=TeX,Scale=1}
\fi
\usepackage{lmodern}
\ifPDFTeX\else
\fi
\IfFileExists{upquote.sty}{\usepackage{upquote}}{}
\IfFileExists{microtype.sty}{
  \usepackage[]{microtype}
  \UseMicrotypeSet[protrusion]{basicmath} 
}{}
\makeatletter
\@ifundefined{KOMAClassName}{
  \IfFileExists{parskip.sty}{%
    \usepackage{parskip}
  }{
    \setlength{\parindent}{0pt}
  \setlength{\parskip}{6pt plus 2pt minus 1pt}}
}{
\KOMAoptions{parskip=half}}
\makeatother
\usepackage{xcolor}
\makeatletter
\ifx\paragraph\undefined\else
\let\oldparagraph\paragraph
\renewcommand{\paragraph}{
  \@ifstar
  \xxxParagraphStar
  \xxxParagraphNoStar
}
\newcommand{\xxxParagraphStar}[1]{\oldparagraph*{#1}\mbox{}}
\newcommand{\xxxParagraphNoStar}[1]{\oldparagraph{#1}\mbox{}}
\fi
\ifx\subparagraph\undefined\else
\let\oldsubparagraph\subparagraph
\renewcommand{\subparagraph}{
  \@ifstar
  \xxxSubParagraphStar
  \xxxSubParagraphNoStar
}
\newcommand{\xxxSubParagraphStar}[1]{\oldsubparagraph*{#1}\mbox{}}
\newcommand{\xxxSubParagraphNoStar}[1]{\oldsubparagraph{#1}\mbox{}}
\fi
\makeatother

\usepackage{longtable,booktabs,array}
\usepackage{calc} 
\usepackage{etoolbox}
\makeatletter
\patchcmd\longtable{\par}{\if@noskipsec\mbox{}\fi\par}{}{}
\makeatother
\IfFileExists{footnotehyper.sty}{\usepackage{footnotehyper}}{\usepackage{footnote}}
\makesavenoteenv{longtable}
\usepackage{graphicx}
\usepackage{hyperref}
\makeatletter
\def\maxwidth{\ifdim\Gin@nat@width>\linewidth\linewidth\else\Gin@nat@width\fi}
\def\maxheight{\ifdim\Gin@nat@height>\textheight\textheight\else\Gin@nat@height\fi}
\makeatother
\setkeys{Gin}{width=\maxwidth,height=\maxheight,keepaspectratio}
\makeatletter
\def\fps@figure{htbp}
\makeatother

\makeatletter
\@ifpackageloaded{caption}{}{\usepackage{caption}}
\AtBeginDocument{%
  \ifdefined\contentsname
  \renewcommand*\contentsname{Table of contents}
  \else
  \newcommand\contentsname{Table of contents}
  \fi
  \ifdefined\listfigurename
  \renewcommand*\listfigurename{List of Figures}
  \else
  \newcommand\listfigurename{List of Figures}
  \fi
  \ifdefined\listtablename
  \renewcommand*\listtablename{List of Tables}
  \else
  \newcommand\listtablename{List of Tables}
  \fi
  \ifdefined\figurename
  \renewcommand*\figurename{Figure}
  \else
  \newcommand\figurename{Figure}
  \fi
  \ifdefined\tablename
  \renewcommand*\tablename{Table}
  \else
  \newcommand\tablename{Table}
  \fi
}
\@ifpackageloaded{float}{}{\usepackage{float}}
\floatstyle{ruled}
\@ifundefined{c@chapter}{\newfloat{codelisting}{h}{lop}}{\newfloat{codelisting}{h}{lop}[chapter]}
\floatname{codelisting}{Listing}

\makeatother
\makeatletter
\@ifpackageloaded{caption}{}{\usepackage{caption}}
\@ifpackageloaded{subcaption}{}{\usepackage{subcaption}}
\makeatother

\ifLuaTeX
\usepackage{selnolig}  
\fi
\usepackage[]{natbib}
\newcommand{\bmtheta}{\bm{\theta}}
\def\bw{{\mathbf{w}}}
\def\bx{{\mathbf{x}}}

\DeclareMathOperator{\sign}{sign}

\newcommand{\anon}{1}

\usepackage{amssymb} 
\usepackage{amsmath} 
\usepackage{amsthm}
\usepackage{makecell}
\usepackage{enumitem}
\usepackage{mathrsfs} 
\usepackage{newtxmath}
\usepackage{listings}
\usepackage{bm}
\usepackage[utf8]{inputenc} 
\usepackage{algorithm, algorithmic}
\usepackage{natbib}
\usepackage{hyperref}
\usepackage{bookmark}
\IfFileExists{xurl.sty}{\usepackage{xurl}}{} 
\hypersetup{
  pdftitle={Title},
  pdfauthor={Author 1; Author 2},
  pdfkeywords={3 to 6 keywords, that do not appear in the title},
  colorlinks=true,
  linkcolor=blue,      
  citecolor=blue,      
  urlcolor=cyan,       
  filecolor=Maroon,
  pdfcreator={LaTeX via pandoc}
}
\usepackage{graphicx}
\usepackage{multirow}
\usepackage{threeparttable}
\usepackage{listings} 
\usepackage{pdfpages}
\usepackage{enumitem}
\usepackage{booktabs} 
\usepackage{textcase}
\newtheorem{theorem}{Theorem}[section]
\newtheorem{lemma}[theorem]{Lemma}

\newtheorem{proposition}[theorem]{Proposition}

\newcommand{\muone}{\mu_1}
\newcommand{\mumone}{\mu_{-1}}
\newcommand{\Dcal}{\mathcal{D}} 
\newcommand{\Xcal}{\mathcal{X}}

\newcommand{\CX}{C_X} 
\newcommand{\CK}{C_{\mathcal K}} 
\newcommand{\norm}[1]{\left\| #1 \right\|} 

\providecommand{\bx}{\boldsymbol{x}}
\newcommand{\bmX}{\bm{X}}

\begin{document}

\def\spacingset#1{\renewcommand{\baselinestretch}%
{#1}\small\normalsize} \spacingset{1}

\if1\anon
{
  \title{\bf Neural Networks of Outcome Weighted Learning for Individualized Treatment Rules}
  \author{Zhu Wang\hspace{.2cm}\\
    Department of Preventive Medicine\\
    University of Tennessee Health Science Center
  }
  \maketitle
} \fi


\if0\anon
{
  \bigskip
  \bigskip
  \bigskip
  \begin{center}
    {\LARGE\bf Neural Networks of Outcome Weighted Learning for Individualized Treatment Rules}

  \end{center}
  \medskip
} \fi

\bigskip
\begin{abstract}
Individualized treatment rules (ITRs) formalize precision medicine by assigning treatments according to patient covariates, with the goal of maximizing expected clinical outcomes. Such rules are especially important when treatment effects vary across patients, as in chronic diseases where demographic, clinical, genetic, imaging, or biomarker information may modify the relative benefits of available therapies. Outcome weighted learning (OWL) estimates ITRs by recasting treatment assignment as a weighted classification problem that directly targets clinical value. Motivated by the flexibility of modern neural networks, we extend single hidden-layer neural-network OWL (NNOWL) from ridge-type regularization to nonlinear variable selection and kernel-based approximation. We establish nonasymptotic convergence rates for these estimators, and study the global convergence and implicit bias of gradient descent for NNOWL. Finally, we extend the neural-network methods from OWL to residual weighted learning. Simulation studies illustrate the roles of overparameterization, kernel approximation, and nonlinear variable selection, and a data application in Alzheimer's disease demonstrates the proposed methods.
\end{abstract}

\noindent%
{\it Keywords:}
Convergence rate;
Global convergency;
Implicit bias;
Kernel method;
Sobolev space;
Variable selection

\vfill

\newpage
\spacingset{1.8} 
\spacingset{1.75} 







\date{}




\section{Introduction}

Personalized medicine seeks treatment rules that respond to patient heterogeneity rather than prescribing the same intervention to everyone. This goal is especially relevant in complex chronic diseases, including neurodegenerative conditions such as Alzheimer's disease, where treatment effects may vary across individuals and evolve over time. In such settings, clinically relevant covariates that may contribute to patient heterogeneity include demographic and clinical characteristics, genetic risk markers, amyloid positron emission tomography (PET) measures, and volumetric magnetic resonance imaging (MRI) measures. These considerations lead naturally to the study of individualized treatment rules (ITRs), which assign treatment as a function of observed covariates with the goal of optimizing clinical outcomes. Broadly speaking, the goal is to learn an optimal treatment rule from data using statistical and machine learning methods \citep{kosorok2019precision}.

One common strategy estimates ITRs through regression-based learning. In particular, $Q$-learning models conditional mean outcomes as functions of patient covariates and treatment, and then derives treatment decisions from the fitted outcome model \citep{qian2011performance}. Neural networks provide a flexible machine learning alternative for estimating such outcome models \citep{james2021introduction, shi2019adapting}.

Outcome weighted learning (OWL) takes a different route and casts ITR estimation as a weighted classification problem \citep{zhao2012estimating}. Instead of fitting a response surface, OWL directly learns treatment assignments by weighting classification errors with the observed clinical reward or outcome. After replacing the discontinuous weighted $0$--$1$ criterion with a surrogate loss, the problem becomes an empirical risk minimization problem that is amenable to computation. \citet{wang2026general} generalized policy calibration theory to broad families of OWL surrogates and obtained corresponding convergence rates for kernel based learning; see also \citet{jiang2024deep} for a summary of related developments, including neural network versions of OWL and its residual weighted counterpart \citep{zhou2017residual, liang2018estimating,bennett2020efficient}. However, the methodology and theoretical properties of neural network OWL remain much less developed.

Several gaps motivate the present work. First, typical neural networks are implemented with ridge type $\ell_2$ regularization, and nonlinear variable selection methods for NNOWL remain underdeveloped. In addition, kernel based methods induced by neural networks can lead to substantial computational advantages, but such methods have not been adapted to NNOWL. Second, convergence rates for these NNOWL based methods are needed to understand how they adapt to low-dimensional structure and support nonlinear variable selection. Third, since the optimization landscape is generally nonconvex, even when the surrogate loss function in NNOWL is convex, whether global convergence can still be obtained for gradient descent remains an important question. Another important aspect of gradient descent is its implicit regularization in NNOWL. Gradient descent often favors low-complexity predictors even when no explicit regularizer is imposed, and understanding this phenomenon has become a central topic in modern machine learning theory, as noted in \citet{james2021introduction}; see \citet{vardi2023implicit} for a general review. Although mean field theory provides a natural framework for analyzing gradient descent in overparameterized neural networks \citep{chizat2018global, chizat2020implicit}, it has not been adapted to NNOWL.

The present work studies single-hidden-layer networks, which are flexible enough to represent nonlinear functions \citep{bishop2006pattern} and can therefore model complex treatment boundaries. The main contributions of this article are as follows. First, on the methodological side, we propose nonlinear variable selection and kernel-based implementations for neural-network OWL (NNOWL) and neural-network residual weighted learning (NNRWL). Second, on the statistical side, we derive nonasymptotic convergence rates for NNOWL with $\ell_2$ and $\ell_1$ penalties, as well as for neural-network-induced kernel methods. We show how neural network classes can adapt to latent low-dimensional structure in the covariates and how large network widths can be used to keep the generalization error under control. Third, on the optimization side, we study gradient descent dynamics in the overparameterized regime through mean-field theory and establish results on global convergence and implicit bias for NNOWL. This analysis characterizes the implicit bias of gradient descent through a max-margin problem with data-dependent geometry. 

The remainder of the article is organized as follows. Section~\ref{sec:est} describes NNOWL with $\ell_2$ and $\ell_1$ regularization, induced kernel methods, and the corresponding extension to NNRWL. Section~\ref{sec:no3} presents the theoretical results for NNOWL, including convergence rates, global convergence, and implicit bias. Section~\ref{sec:sim} reports the simulation studies, and Section~\ref{sec:data} presents an application to a randomized trial of asymptomatic individuals at risk for Alzheimer's disease \citep{sperling2023trial}. Section~\ref{sec:dis} concludes with a discussion. Proofs and detailed analysis of global convergence and implicit bias of NNOWL are collected in the Supplementary Material.

\section{Methodology}\label{sec:est}
We begin by reviewing formulations of individualized treatment rule estimation as a weighted classification problem, including neural network decision rules for OWL and their extension to residual weighted learning. We then propose regularized variants that encourage sparse solutions, along with related kernel based implementations for comparison.
\subsection{Neural-Network Formulations of OWL and RWL}\label{sec:nnowl}
We focus on a two-arm randomized trial with observations $(\bmX,A,R)$, where $\bmX\in\Xcal\subset\mathbb{R}^d$ denotes the baseline covariates, $A\in\{1,-1\}$ denotes treatment assignment, and $R\ge 0$ is the observed reward or outcome with larger values preferred. A decision rule is a measurable map $\Dcal:\Xcal\to\{-1,1\}$. For such a rule, the corresponding population value is
\(
  V(\Dcal) = \mathbb{E}[R \mid A = \Dcal(\bmX)] = \mathbb{E}\left[\frac{R}{\pi(A,\bmX)} \mathbb{I}(A = \Dcal(\bmX))\right],
\)
where $\pi(A,\bmX)$ is the propensity score evaluated at the observed treatment and covariates, with $\pi(a,\bx)=P(A=a\mid \bmX=\bx)$, and $\mathbb{I}(\cdot)$ is the indicator function \citep{qian2011performance, zhao2012estimating}. In a randomized trial, this propensity is known from the experimental design. The statistical task is to estimate a rule $\Dcal$ that maximizes $V(\Dcal)$.

Unlike deep neural networks for OWL and RWL \citep{liang2018estimating, jiang2024deep}, we use a single hidden-layer neural network that maps an input vector $\bx \in \mathbb{R}^d$ to a prediction by combining $m$ elementary building blocks, or neurons. Each neuron applies an activation function $\sigma$ to an affine transformation of the input, and the network output is the sum of these contributions:
\(
  f(\bx) \;=\; \sum_{j=1}^{m} \zeta_j \,\sigma\!\left(\bw_j^\top \bx + b_j\right).
\)
Here $\bw_j \in \mathbb{R}^d$ and $b_j \in \mathbb{R}$ are the input weights and bias for neuron $j$, while $\zeta_j \in \mathbb{R}$ is its output weight. Together, these parameters determine each neuron's response to the input and its contribution to the final predictor.
The activation function $\sigma$ introduces the nonlinearity that makes neural networks substantially more expressive than linear models. A common choice is the rectified linear unit (ReLU):
\(
  \sigma(u) \;=\; (u)_+ \;=\; \max\{u,\,0\}.
\)

Let $\{(\bx_i,a_i,r_i)\}_{i=1}^n$ be independent and identically distributed observations. Estimating the network parameters
$\bmtheta = \big((\zeta_j,\bw_j,b_j)\big)_{j=1}^m$ determines the prediction
$f_{\bmtheta}(\bx)=\sum_{j=1}^{m} \zeta_j\,\sigma\!\left(\bw_j^\top \bx + b_j\right)$ and the induced decision rule $\operatorname{sign}(f_{\bmtheta}(\bx))$. We call this neural-network outcome weighted learning (NNOWL). A direct NNOWL formulation therefore minimizes the weighted empirical $0$--$1$ loss
\(
  \frac{1}{n} \sum_{i=1}^{n}
  \frac{r_{i}}{\pi\left(a_{i}, \bx_{i}\right)}
  \mathbb{I}\!\left(a_i \neq \operatorname{sign}(f_{\bmtheta}(\bx_i))\right).
\)
As is typical in machine learning, we replace the $0$--$1$ loss with a surrogate loss $\ell$ and add a standard ridge penalty, yielding NNOWL-Ridge:
\begin{equation}
  \label{eq:erm}
  \min_{\bmtheta \in \mathbb{R}^{m(d+2)}} \;\frac{1}{n} \sum_{i=1}^{n}
  \frac{r_{i}}{\pi\left(a_{i}, \bx_{i}\right)}
  \ell\!\left(a_i \sum_{j=1}^{m} \zeta_j\,\sigma\!\left(\bw_j^\top \bx_i + b_j\right)\right)
  +\frac{\lambda}{2}\sum_{j=1}^{m}\left(\zeta_j^2+\|\bw_j\|_2^2+b_j^2\right),
  \qquad \lambda\ge 0.
\end{equation}
Numerical solutions are typically obtained by gradient descent or stochastic gradient descent \citep{james2021introduction}.


RWL reduces outcome variability and is invariant to shifts and positive rescalings of the reward by replacing the observed reward $r_i$ with a residual $\hat{r}_i$ after removing prognostic main effects. Once the main-effect function has been estimated and residuals $\hat r_i$ have been formed, RWL can be expressed in the same weighted-classification form as OWL by replacing the original treatment label and reward weight with
\(
  \tilde y_i=a_i\operatorname{sign}(\hat r_i), 
  W_i=\frac{|\hat r_i|}{\pi(a_i,\bx_i)}.
\)
Thus, NNRWL uses the same neural-network score $f_{\bmtheta}$ and surrogate-loss construction as NNOWL, but replaces $(a_i,r_i/\pi(a_i,\bx_i))$ with $(\tilde y_i,W_i)$. Adding the same ridge penalty gives the NNRWL-Ridge objective:
\begin{equation}
  \min_{\bmtheta \in \mathbb{R}^{m(d+2)}} \;\frac{1}{n} \sum_{i=1}^{n} W_i
  \ell\!\left(\tilde y_i \sum_{j=1}^{m} \zeta_j\,\sigma\!\left(\bw_j^\top \bx_i + b_j\right)\right)
  +\frac{\lambda}{2}\sum_{j=1}^{m}\left(\zeta_j^2+\|\bw_j\|_2^2+b_j^2\right),
  \qquad \lambda\ge 0.
  \label{eq:ermrwl}
\end{equation}

\subsection{Nonlinear Variable Selection for NNOWL and NNRWL}\label{sec:nvs}

When only a subset of the input covariates is active, variable selection can improve interpretability and learning accuracy. We replace the $\ell_2$ regularization in the NNOWL and NNRWL objectives \eqref{eq:erm} and \eqref{eq:ermrwl} with an $\ell_1$ penalty on the first-layer input-weight block $\bw$, leaving the output weights $\zeta$ and biases $b$ unpenalized. The resulting objective has the composite form
\(
  \mathcal{L}(\bmtheta)+\lambda\|\bw\|_1,
\), where $\mathcal{L}$ denotes the corresponding smooth surrogate-loss component. The $\|\bw\|_1$ penalty encourages sparsity in the input weights and thereby supports nonlinear variable selection. We refer to the resulting estimators as NNOWL-VS and NNRWL-VS, respectively.

Compared with the earlier $\ell_2$ penalty, the $\ell_1$ penalty is nonsmooth at zero, so ordinary gradient descent cannot be applied directly to the full objective. We therefore optimize this composite objective by applying proximal gradient to the $\bw$-block while updating the output weights $\zeta$ and hidden-layer biases $b$ by ordinary gradient steps. This modification preserves the smooth-gradient structure of the $\mathcal{L}$ term, rather than treating the entire objective as nonsmooth and relying on the slower convergence guarantees of generic subgradient methods \citep{bach2024learning}.

In each iteration, the unpenalized parameters $\zeta$ and $b$ are updated by ordinary gradient descent on the smooth loss $\mathcal{L}$, whereas the penalized input-weight block $\bw$ is updated by a proximal step. This separation motivates the two step sizes used in the implementation: the usual learning rate $\gamma$ controls the gradient updates for the unpenalized network parameters, while the proximal step size $\gamma_\bw$ controls the update of $\bw$ and hence the soft-thresholding amount for sparsity. With $(\zeta^{(t)},b^{(t)})$ fixed for this part of the iteration, proximal gradient updates the input-weight block by minimizing a local quadratic surrogate of the composite objective:
\[
  \bw^{(t+1)} = \arg\min_{\bw}\left\{
    \begin{aligned}
      &\mathcal{L}\bigl(\zeta^{(t)},\bw^{(t)},b^{(t)}\bigr)
      + \bigl\langle
      \nabla_\bw \mathcal{L}\bigl(\zeta^{(t)},\bw^{(t)},b^{(t)}\bigr),
      \bw-\bw^{(t)}
      \bigr\rangle \\
      &\qquad
      + \frac{1}{2\gamma_\bw}\|\bw-\bw^{(t)}\|_2^2
      + \lambda\|\bw\|_1
    \end{aligned}
  \right\},
\]
where $\gamma_\bw>0$ denotes the step size for updating $\bw$. This surrogate is standard when $\mathcal{L}$ is differentiable in $\bw$ and $\nabla_\bw \mathcal{L}$ is Lipschitz continuous. Completing the square gives the equivalent proximal subproblem
\(
  \bw^{(t+1)} = \arg\min_{\bw}\left\{\frac{1}{2\gamma_bw}\|\bw-z^{(t)}\|_2^2 + \lambda\|\bw\|_1\right\},
\)
where
\(
  z^{(t)} = \bw^{(t)} - \gamma_\bw\nabla_\bw \mathcal{L}\bigl(\zeta^{(t)},\bw^{(t)},b^{(t)}\bigr).
\)
This subproblem has the closed-form soft-thresholding solution
\(
  \bigl[\bw^{(t+1)}\bigr]_j = \mathrm{sign}\bigl(z^{(t)}_j\bigr)\max\bigl(|z^{(t)}_j| - \gamma_\bw\lambda,\, 0\bigr).
\)
Here $z^{(t)}_j$ denotes the $j$th coordinate of the intermediate vector $z^{(t)}$. Thus, each iteration combines ordinary gradient steps for the unpenalized blocks $\zeta$ and $b$ with a proximal-gradient step for the penalized block $\bw$. This update exploits the smooth structure of $\mathcal{L}$ while handling the nonsmooth $\ell_1$ term exactly. In particular, coordinates with magnitude below $\gamma_\bw\lambda$ are set exactly to zero, which is why proximal gradient is well suited to exact variable selection in the input-weight block $\bw$.

\subsection{Implicit Kernel Methods Induced by NNOWL and NNRWL}\label{sec:kernel}

Neural networks are closely related to kernel methods through the Hilbert-space structure induced by their integral representations. Fix a probability measure $\rho_0$ with full support on a parameter set $\mathcal K\subset\mathbb R^{d+1}$, and consider functions on a bounded covariate domain $\mathcal X$ of the form
\(
  f(\bx)=\int_{\mathcal K} (\bw^\top \bx+b)_+\, g(\bw,b)\, d\rho_0(\bw,b),
\)
where $g\in L^2(\rho_0)$ is the density of an absolutely continuous representing measure. Equivalently, this is the Hilbertian $V_2$ construction:
\begin{equation}
  \label{eqn:hilbertian_norm}
  V_2(f):=\inf\left\{\|g\|_{L^2(\rho_0)}:
  f(\bx)=\int_{\mathcal K} (\bw^\top \bx+b)_+g(\bw,b)\,d\rho_0(\bw,b),\ \bx\in\mathcal X\right\}.
\end{equation}
The resulting space $\mathcal F_2=\{f:V_2(f)<\infty\}$ is a reproducing kernel Hilbert space (RKHS) with kernel $\mathsf{k}$ given by \citep[Proposition~9.3]{bach2024learning}: 
\begin{equation}\label{eqn:kernel}
  \mathsf{k}(\bx,\bx')=\int_{\mathcal K} (\bw^\top \bx+b)_+(\bw^\top \bx'+b)_+\, d\rho_0(\bw,b).
\end{equation}

Given training data $\{(r_i,\bx_i,a_i)\}_{i=1}^n$, define the inverse-probability weights $\eta_i=r_i/\pi(a_i,\bx_i)$. We refer to the following regularized empirical OWL problem as the NNOWL-Kernel estimator: 
\begin{equation}
  \label{eq:kernel-owl-functional}
  \min_{f\in\mathcal F_2}
  \left\{
    \frac{1}{n}\sum_{i=1}^n \eta_i\,\ell\bigl(a_i f(\bx_i)\bigr)
    +\frac{\lambda}{2}V_2(f)^2
  \right\}.
\end{equation}

There are two practical ways to construct the kernel matrix used in computation. First, when the integral kernel in \eqref{eqn:kernel} has a closed form, including for certain ReLU feature distributions \citep{bach2024learning}, one may form the exact kernel matrix from all pairwise evaluations $\mathsf{k}(\bx_i,\bx_j)$. This avoids Monte Carlo error but requires storing and manipulating an $n\times n$ matrix. Second, when no closed form is available or when a lower-dimensional explicit representation is computationally preferable, one may use random features: sample hidden parameters $(\bw_j,b_j)_{j=1}^m$ independently from $\rho_0$ and keep them fixed, giving the empirical kernel
\(
  \hat{\mathsf{k}}_m(\bx,\bx')=\frac{1}{m}\sum_{j=1}^m (\bw_j^\top \bx+b_j)_+(\bw_j^\top \bx'+b_j)_+,
\) 
which approximates \eqref{eqn:kernel}. 

Let $K_m$ denote the kernel matrix used in computation, with $K_m=K$ for the exact kernel and $K_m=\widehat K_m$ for the random-feature approximation. By the representer theorem, the minimizer has a finite kernel expansion $f(\cdot)=\sum_{i=1}^n \chi_i \mathsf{k}(\bx_i,\cdot)$. In this representation, $(K_m\chi)_i=f(\bx_i)$ is the fitted score at the $i$th training point, and the penalty satisfies $V_2(f)^2=\chi^\top K_m\chi$. Thus, \eqref{eq:kernel-owl-functional} reduces to the finite-dimensional coefficient problem
\begin{equation}
  \label{eq:kernel-owl-objective}
  \min_{\chi\in\mathbb{R}^n}
  \left\{
    \frac{1}{n}\sum_{i=1}^n \eta_i\,\ell\bigl(a_i (K_m\chi)_i\bigr)
    +\frac{\lambda}{2}\chi^\top K_m \chi
  \right\}.
\end{equation}
Once the kernel matrix is fixed, estimation therefore reduces to optimizing the output coefficients. Regularization properties and computational algorithms for this kernel-based OWL formulation can then be obtained as in \citet{wang2026general}.

The same kernel estimator can be adapted to NNRWL by applying the residual-weighting transformation to the OWL objective in \eqref{eq:kernel-owl-functional}. After estimating the main-effect model and forming residuals $\hat r_i$, we replace the OWL label $a_i$ with the pseudo-label $\tilde y_i=a_i\operatorname{sign}(\hat r_i)$ and replace the inverse-probability weight $\eta_i=r_i/\pi(a_i,\bx_i)$ with the nonnegative residual weight $W_i=|\hat r_i|/\pi(a_i,\bx_i)$. The NNRWL-Kernel estimator therefore minimizes the same regularized RKHS objective, with the loss term $W_i\ell\{\tilde y_i f(\bx_i)\}$ in place of $\eta_i\ell\{a_i f(\bx_i)\}$.

The RKHS $\mathcal F_2$ induced by the fixed hidden-layer distribution differs from the variation-norm viewpoint for fully trained two-layer networks. Because the hidden-layer distribution fixes the RKHS $\mathcal F_2$, kernel learning estimates functions in this Hilbert space rather than optimizing hidden parameters jointly with output weights. As shown by \citet{bach2017breaking}, the RKHS class is contained in the larger class obtained by training all network weights: the variation norm $V_1$ allows less regular functions, including individual ReLU neurons, whereas the Hilbertian $V_2$ norm gives a more restrictive and smoother class. Thus, fixed-kernel methods often produce smoother fits, while fully trained neural networks can adapt more locally to heterogeneous structure. The RKHS approach also offers computational advantages, including simpler optimization and easier hyperparameter tuning than gradient-descent training of full neural networks, as illustrated in the numerical studies below.

\section{Theoretical Results}\label{sec:no3}

This section studies the convergence rates of NNOWL estimators under various penalized settings. We also investigate properties of the gradient-descent algorithm used for optimization.
Let $\mathcal R(f)$ denote the weighted $0$--$1$ risk with Bayes value $\mathcal R^{*}$, and let $\mathcal R_\ell(f)$ denote the surrogate risk with optimum $\mathcal R_{\ell}^{*}$.
More explicitly,
\(\mathcal{R}(f)=\mathbb{E}\left[\frac{R}{\pi(A,\bmX)}\mathbb{I}\{A\neq\operatorname{sign}(f(\bmX))\}\right]\) and
\(\mathcal{R}_\ell(f)=\mathbb{E}\left[\frac{R}{\pi(A,\bmX)}\ell(Af(\bmX))\right]\). Assuming no ties, the optimal ITR is $\Dcal^*(\bx)=\operatorname{sign}\{\mu_1(\bx)-\mu_{-1}(\bx)\}$, where $\mu_a(\bx)=\mathbb{E}[R\mid\bmX=\bx,A=a]$. When the surrogate optimum is attained, let $f_*$ satisfy $\mathcal R_{\ell}^{*}=\mathcal R_\ell(f_*)$. Consider an empirical surrogate-risk minimizer $\hat f_D$ over a constrained class $\mathcal F_D$.
The target risk is controlled through a calibration comparison of the form $\mathcal R(f)-\mathcal R^{*}\leq\Upsilon\{\mathcal R_\ell(f)-\mathcal R_{\ell}^{*}\}$ for an appropriate transform $\Upsilon$ and all $f\in\mathcal F_D$, in particular for $\hat f_D$.
The excess surrogate risk is then decomposed as
\[
  \mathcal R_\ell(\hat f_D)-\mathcal R_{\ell}^{*}
  =\{\mathcal R_\ell(\hat f_D)-\inf_{f\in\mathcal F_D}\mathcal R_\ell(f)\}
  +\{\inf_{f\in\mathcal F_D}\mathcal R_\ell(f)-\mathcal R_{\ell}^{*}\},
\]
where the two terms are the estimation and approximation errors, respectively. Together with the calibration transform $\Upsilon$, this decomposition reduces control of the excess $0$--$1$ risk to bounding these two components of the excess surrogate risk. We now describe the calibration bound and the two components in detail.

\subsection{Excess Risk Upper Bound}

Many surrogate loss functions can be used for NNOWL, and the calibration transform $\Upsilon$ determines how the corresponding excess surrogate risk upper bounds the excess $0$--$1$ risk \citep{wang2026general}. In this article, we focus on the logistic loss $s(u)=\log(1+\exp(-u))$; we also consider the exponential loss $s(u)=\exp(-u)$ because its exponential tail leads to the same implicit-bias phenomenon discussed in Section~\ref{sec:global}. The following lemma gives the form of $\Upsilon$ used for both losses. 
\begin{lemma}
  \label{lem:excess_risk_bound}
  Assume that $\muone+\mumone\le M$ for some $M>0$.
  For the logistic and exponential losses, the excess $0$--$1$ risk satisfies
  \(
    \mathcal{R}(f)-\mathcal{R}^{*}
    \le
    \sqrt{2M \bigl(\mathcal{R}_{\ell}(f)-\mathcal{R}_{\ell}^{*}\bigr)}.
  \)
\end{lemma}
Thus, for the logistic and exponential losses considered here, the calibration transform $\Upsilon(t)=\sqrt{2Mt}$ converts an excess surrogate-risk bound of order $t$ into an excess $0$--$1$ risk bound of square-root order. Other losses may yield sharper calibration transforms, as discussed in \citet{wang2026general}.

\subsection{Approximation Error}

Approximation error measures how well the neural-network class approximates the population target function $f_*$. For ReLU networks, the estimation error derived below scales with the output-weight $\ell_1$ norm, not directly with the number of hidden neurons. Thus, the approximation analysis should control the infinite-width analogue of this $\ell_1$ norm rather than network width.

We use the variation-norm framework of \citet{bach2024learning} to formalize this idea. Let
\(
  \mathcal K=\left\{(\bw,b)\in\mathbb{R}^{d+1}:\|\bw\|_{2}=1,\ |b| \leqslant \CK\right\}.
\)
This compact set indexes normalized ReLU neurons. A finite-width network
\(f(\bx)=\sum_{j=1}^m \zeta_j(\bw_j^\top\bx+b_j)_+\)
can be written as an integral by taking the signed measure
\(\nu=\sum_{j=1}^m \zeta_j\delta_{(\bw_j,b_j)}\), in which case
\(\int_{\mathcal K}|d\nu|=\|\zeta\|_1\). Replacing this finitely supported measure by an arbitrary signed Radon measure gives the infinite-width closure of the same model class while preserving the same complexity measure.
For functions admitting such an integral representation with a signed Radon measure $\nu$ on $\mathcal K$, define
\begin{equation}
  \label{eqn:variation_norm}
  V_1(f) = \inf_{\nu \in \mathcal{M}(\mathcal K)} \left\{ \int_{\mathcal K} |d\nu(\bw, b)| : f(\bx) = \int_{\mathcal K} (\bw^\top \bx + b)_+ \, d\nu(\bw, b) \right\},
\end{equation}
where $\mathcal{M}(\mathcal K)$ is the set of signed measures on $\mathcal K$ with finite total variation. The quantity $V_1(f)$ measures the complexity of $f$ through its representation as an integral over ReLU neurons, and it is the natural infinite-width analogue of an $\ell_1$ penalty on the output weights. Let
\(
  \mathcal F_1=\{f:V_1(f)<\infty\}
\)
denote the corresponding variation-norm space. This construction yields a normed linear space of functions generated by bounded-variation mixtures of neurons; moreover, since the neuron dictionary $\mathcal K$ is compact, functions with finite variation norm are Lipschitz on bounded covariate domains. Positive homogeneity of the ReLU activation also makes the normalization of this dictionary mainly a matter of convention: scaling can be shifted between hidden-unit parameters and output weights without changing the represented function.

For approximation, we need conditions under which the target $f_*$, or a good approximation to it, has finite and preferably small $V_1$ norm. To connect this requirement to standard smoothness assumptions, let $H^s(\mathbb{R}^d)$ denote the Sobolev space of order $s$, with norm
\begin{equation}\label{eq:sobolev}
  \|f\|_{H^s}^2 = \int_{\mathbb{R}^d} (1 + \|\omega\|_2^2)^s |\hat{f}(\omega)|^2 \, d\omega \leq C_s < \infty,
\end{equation}
where $\hat{f}$ denotes the Fourier transform of $f$. The following lemma states the key link between Sobolev smoothness and variation-norm complexity:
\begin{lemma}
  \label{lem:gamma1_upper}
  If $s=d/2+5/2$ and $f\in H^s(\mathbb{R}^d)$, then
  \(
    V_1(f)\le C_1\|f\|_{H^s},
  \)
  where $C_1>0$ depends only on the dictionary radius $\CK$ and the dimension $d$.
\end{lemma}
The Sobolev approximation bound below, adapted from \citet[Lemma~3.1]{wang2026general}, is then used to quantify the approximation error.
\begin{lemma}
  \label{lem:matern}
  Assume that $\ell$ is $G$-Lipschitz continuous, $\frac{R}{\pi(A, \bmX)}\leq B$ almost surely, $f_* \in H^t(\mathbb{R}^d)$,
  that $P_X$ admits a bounded Lebesgue density $p$, and that $\varsigma > 0$. Let $C_s$ be an upper bound given in \eqref{eq:sobolev}, and let $C_k>0$ denote a Fourier-normalization constant. Then
  \begin{equation*}
    \inf_{f \in H^s} \left( \mathcal{R}_\ell(f) - \mathcal{R}_\ell^* + \frac{\varsigma}{2} \|f\|_{H^s} \right)
    \leq
    \begin{cases}
      \frac{\varsigma}{\sqrt{2}}\|f_*\|_{H^s}  &\text{ if } t \geq s\\
      \varsigma^{t/s}
      2^{1/2-t/s}
      (BG)^{1-t/s}
      \sqrt{
        \frac{
      C_s \|p\|_\infty^{1-t/s} }{(2\pi)^d C_k^{t/s}}} &\text{ if } t < s.
    \end{cases}
  \end{equation*}
\end{lemma}

\subsection{Convergence Rates for NNOWL-Ridge}
We next derive the convergence rate for the ridge-regularized NNOWL estimator adapting \citet{bach2024learning}. The analysis utilizes the positive homogeneity of the ReLU activation. For any $\beta>0$, $\sigma(\beta u)=\beta\sigma(u)$, so the single-neuron contribution
\(
  \zeta_j\sigma(\bw_j^\top\bx+b_j)
\)
can equivalently be written as
\(
  (\beta_j\zeta_j)\sigma((\bw_j^\top\bx+b_j)/\beta_j).
\)
Thus, the represented function is unchanged if the output weight is multiplied by $\beta_j$ and the input weights and bias are divided by $\beta_j$. However, the ridge penalty changes as the per-neuron squared $\ell_2$ penalty becomes
\(
  \beta_j^2\zeta_j^2 + \beta_j^{-2}\left(\|\bw_j\|_2^2 + b_j^2/\CK^2\right).
\)
Minimizing this expression over the free scale $\beta_j>0$ gives
\(
  2\,|\zeta_j|\left(\|\bw_j\|_2^2 + b_j^2/\CK^2\right)^{1/2},
\)
which is an $\ell_1$-type penalty on the output weight multiplied by the norm of the hidden-unit parameters.
Thus, once hidden-unit parameters are normalized, the ridge penalty induces an $\ell_1$-type control on the output weights. This is the same complexity measure used in the variation-norm formulation above.

Although NNOWL-Ridge is implemented through the penalized objective \eqref{eq:erm}, the rate analysis is simpler when stated for a constrained class. Motivated by the preceding rescaling argument, we use the canonical normalization
$\|\bw_j\|_2^2 + b_j^2/\CK^2 = 1$ for each neuron and impose the output-weight constraint $\|\zeta\|_1 \leq D$. This gives the feasible parameter set
\[
\Theta^{(2)}=\left\{\bmtheta:\|\bw_j\|_2^2+b_j^2/\CK^2=1,\ j=1,\dots,m,\ \|\zeta\|_1\le D\right\}.
\]
Constraining the network parameters to lie in $\Theta^{(2)}$, we obtain the finite-width class
\(
  \mathcal{F}^{(2)}_{m,D} \;=\; \bigl\{\,f_{\bmtheta} : \bmtheta \in \Theta^{(2)}\,\bigr\}.
\)
Using concentration inequalities and Rademacher-complexity bounds for this class, the estimation error can be upper bounded by a quantity that does not depend explicitly on the nominal width $m$, as summarized in the following theorem.

\begin{theorem}\label{thm:estimation_error}
Assume that the loss function is $G$-Lipschitz continuous and that $\|\bmX\|_2 \leq \CX$ and $\frac{R}{\pi(A, \bmX)} \leq B$ almost surely. If $\hat{f}^{(2)}_{m,D}$ is an empirical risk minimizer over $\mathcal{F}^{(2)}_{m,D}$, then
  $$
  \mathbb{E}\left[\mathcal{R}_\ell(\hat{f}^{(2)}_{m,D})-\inf _{f \in \mathcal{F}^{(2)}_{m,D}} \mathcal{R}_\ell(f)\right] \leqslant \frac{8 G D (\CX+\CK) B}{\sqrt{n}}.
  $$
\end{theorem}

We now combine the approximation error with
the estimation bound from Theorem~\ref{thm:estimation_error}.

\begin{theorem}
  \label{thm:balanced_approximation_bound}
  Assume the conditions of Theorem~\ref{thm:estimation_error} hold, with $\hat f_D$ minimizing the empirical surrogate risk over $\{f\in\mathcal F_1:V_1(f)\le D\}$. Then:
  \begin{enumerate}
    \item
      \(
        \inf_{D\ge 0}
        \mathbb E\left[\mathcal R_\ell(\hat f_D)-\mathcal R_{\ell}^{*}\right]
        \leq \inf_{f\in\mathcal F_1}
        \left\{
          \frac{8\,G(\CX+\CK)B}{\sqrt n}V_1(f)
          +
          \left(\mathcal{R}_\ell(f)-\mathcal R_\ell^*\right)
        \right\}.
      \)
    \item
     For an optimal $D \le \frac{\sqrt{n}}{8 G B (\CX+\CK) \sqrt{2}} \varepsilon_n^{(2)}$, we have
     \(
        \mathbb E\left[\mathcal R_\ell(\hat f_D)-\mathcal R_{\ell}^{*}\right]
        \le
        \varepsilon_n^{(2)},
      \)
      where
      \[
        \varepsilon_n^{(2)}
        =
        GB\sqrt{
          2\inf_{f\in\mathcal F_1}
          \left\{
            \|f-f_*\|_{L_2(P_X)}^2+\frac{64(\CX+\CK)^2}{n}\{V_1(f)\}^2
          \right\}
        }.
      \]
  \end{enumerate}
\end{theorem}
The main value of Theorem~\ref{thm:balanced_approximation_bound} is that it converts approximation properties of the target function $f_*$ into convergence-rate bounds. We first state the consequence under ordinary Sobolev smoothness: when no additional structure is assumed, smoothness alone yields a valid nonparametric rate, but its exponent depends on the ambient dimension $d$.

\begin{proposition}
  \label{prop:risk_rate}
  Assume that the conditions of Theorem~\ref{thm:balanced_approximation_bound} hold. Suppose that $f_* \in H^t$, and let $s=d/2+5/2$. Then, for a suitable choice of the radius $D$, there exist constants
  $\tilde{C} > 0$ and $\bar C_{t,d}>0$ such that
  \begin{equation*}
    \mathbb E\left[\mathcal R_\ell(\hat f_D)-\mathcal R_{\ell}^{*}\right]
    \leq
    \begin{cases}
      8\,G\tilde{C}B\,\|f_*\|_{H^s}n^{-1/2}
      & \text{if } t \geq s,\\[1.2ex]
      \bar C_{t,d}\,n^{-t/(d+5)} & \text{if } t < s.
    \end{cases}
  \end{equation*}
  where $\tilde C$ depends only on the covariate radius $\CX$, the dictionary radius $\CK$, and $d$, and $\bar C_{t,d}$ depends on $t$, $d$,
  $G$, $B$, $\CX$, $\CK$, $\|p\|_\infty$, and the constants $C_s$ and $C_k$ from Lemma~\ref{lem:matern}.
\end{proposition}
If $f_*$ is sufficiently smooth to belong to a Sobolev space of order $t \geq d/2+5/2$, then the parametric rate $n^{-1/2}$ is attained. In the more typical nonsmooth case, domain and extension conditions ensuring that a Lipschitz $f_*$ belongs to a first-order Sobolev class allow us to take $t=1$, giving
\(
  \mathbb E\left[\mathcal R_\ell(\hat f_D)-\mathcal R_{\ell}^{*}\right]
  =
  O\bigl(n^{-1/(d+5)}\bigr).
\)
This rate explicitly exhibits the curse of dimensionality: when smoothness is imposed on the full $d$-dimensional covariate space, the exponent deteriorates as $d$ grows. The rate is also conservative, since the Sobolev comparison used here takes $s=d/2+5/2$, whereas sharper analyses can yield $s=d/2+3/2$ and hence replace $d+5$ by $d+3$ in the nonparametric rate \citep{bach2017breaking,bach2024learning}.

The important advantage of the variation norm $V_1$ is that the preceding approximation--estimation tradeoff can be applied to lower-dimensional or simpler representations when they exist. The next proposition records two cases. The first item treats a latent linear subspace, where the intrinsic dimension $r$ replaces the ambient dimension $d$ in the Sobolev comparison. The second item treats a finite teacher-network representation, where the variation norm is controlled directly by the number and weights of the representing neurons.
\begin{proposition}
  \label{prop:adaptivity}
  Assume that the conditions of Theorem~\ref{thm:balanced_approximation_bound} hold. Then, for a suitable choice of the radius $D$, the following rates hold.
  \begin{enumerate}
    \item Suppose that $f_*(\bx)=g(\mathbf{V}^\top \bx)$ depends only on an unknown $r$-dimensional subspace, where $r < d$, and that $g \in H^t$. Then the rates in Proposition~\ref{prop:risk_rate} hold with $s=r/2 + 5/2$, and the associated constants are independent of the ambient dimension $d$.
    \item If $f_*$ is a linear combination of $k$ hidden neurons, then
      \(
        \mathbb E\left[\mathcal R_\ell(\hat f_D)-\mathcal R_{\ell}^{*}\right]
        \leq k\tilde{C}_k/\sqrt n,
      \)
      where $\tilde{C}_k$ depends on $G$, $\CX$, $\CK$, $B$, and the weights assigned to the $k$ hidden neurons in the representation of $f_*$.
  \end{enumerate}
\end{proposition}
In particular, when $g$ is Lipschitz in the latent-subspace setting, the rate becomes $O(n^{-1/(r+5)})$, replacing the ambient dimension by the latent dimension. In the teacher-network setting, the bound avoids the ambient-dimensional Sobolev exponent altogether, up to the representation size $k$.

\subsection{Convergence Rates for NNOWL-VS}
For NNOWL-VS, the computational estimator was introduced in Section~\ref{sec:nvs}. Here we state its rate analysis using a constrained class, replacing the previous $\ell_2$ first-layer normalization by an $\ell_1$ constraint on the input-affine parameters. The bias term is included in this theoretical normalization for analytical convenience, so that each hidden unit has a bounded normalized affine argument.
The corresponding finite-width parameter set and network class are
\(\Theta^{(1)}=\{\bmtheta:\|\bw_j\|_1+|b_j|/\CK=1,\ j=1,\dots,m,\ \|\zeta\|_1\le D\}\)
and \(\mathcal F^{(1)}_{m,D}=\{f_{\bmtheta}:\bmtheta\in\Theta^{(1)}\}\).
We denote the associated infinite-width variation-norm ball by $\mathcal F_D^{(1)}$.
\begin{theorem}
  \label{thm:nonlinear_variable_selection}
  Assume that $\ell$ is $G$-Lipschitz, that $\|\bmX\|_\infty \le \CX$ almost surely, and
  that $R/\pi(A,\bmX)\le B$ almost surely. If $\hat f^{(1)}_{m,D}$ is an empirical surrogate-risk
  minimizer over $\mathcal F^{(1)}_{m,D}$, then
  \(
    \mathbb{E}\left[\mathcal{R}_\ell(\hat{f}^{(1)}_{m,D})-\inf _{f \in \mathcal{F}^{(1)}_{m,D}} \mathcal{R}_\ell(f)\right]
    \le
    8\sqrt{2}\,GB(\CX+\CK)D\sqrt{\frac{\log(2(d+1))}{n}}.
  \)
\end{theorem}

Compared with the NNOWL-Ridge estimation bound in Theorem~\ref{thm:estimation_error}, the NNOWL-VS bound has the same $D/\sqrt n$ dependence but includes the additional factor $\sqrt{\log(2(d+1))}$ from the $\ell_1$ input constraint and $\ell_\infty$ covariate control. This logarithmic dimension dependence is the price paid for nonlinear variable-selection structure. We now combine this estimation bound with the approximation argument exactly as in the $\ell_2$ analysis above.

\begin{theorem}
  \label{thm:nonlinear_variable_selection_balance}
  Assume the conditions of Theorem~\ref{thm:nonlinear_variable_selection} hold, with $\hat f_D^{(1)}$ minimizing the empirical surrogate risk over $\mathcal F_D^{(1)}$. Then:
  \begin{enumerate}
    \item
      \(
        \inf_{D\ge 0}
        \mathbb E\left[\mathcal R_\ell(\hat f_D^{(1)})-\mathcal R_\ell^*\right]
        \leq
        \inf_{f\in\mathcal F_1}
        \left\{
          8\sqrt{2}\,GB(\CX+\CK)\sqrt{\frac{\log(2(d+1))}{n}}\,V_1(f)
          +
          \left(\mathcal{R}_\ell(f)-\mathcal R_\ell^*\right)
        \right\}.
      \)
    \item
      For an optimal
      $D \le \frac{\sqrt n}{16GB(\CX+\CK)\sqrt{\log(2(d+1))}}\varepsilon_n^{(1)}$, 
      we have
      \(\mathbb E[\mathcal R_\ell(\hat f_D^{(1)})-\mathcal R_\ell^*]
      \le \varepsilon_n^{(1)}\),
      where
      \[
        \varepsilon_n^{(1)}
        =
        GB\sqrt{
          2\inf_{f\in\mathcal F_1}
          \left\{
            \|f-f_*\|_{L_2(P_X)}^2
            + \frac{128(\CX+\CK)^2\log(2(d+1))}{n}\{V_1(f)\}^2
          \right\}
        }.
      \]
  \end{enumerate}
\end{theorem}
The following proposition shows that this bound yields a sharper rate when the target depends only on a sparse subset of predictors.
\begin{proposition}
  \label{prop:nonlinear_variable_selection_rate}
  Assume the conditions of Theorem~\ref{thm:nonlinear_variable_selection_balance} hold.
  Assume that there exists a subset $J\subset\{1,\dots,d\}$ with $|J|=k$ and a function $g\in H^t(\mathbb R^k)$ such that $f_*(\bx)=g(\bx_J)$. Let $s=k/2+5/2$. Then, for a suitable choice of the radius $D$, there exists a
  constant $\tilde C_k>0$, depending only on $\CX$, $\CK$, and $k$, such that
  \[
    \mathbb E\left[\mathcal R_\ell(\hat f_D^{(1)})-\mathcal R_\ell^*\right]
    \le
    \begin{cases}
      8\sqrt{2}\,G\tilde C_k B\,\|g\|_{H^s}\sqrt{\dfrac{\log(2(d+1))}{n}}, & \text{if } t\ge s,\\[1.2ex]
      \bar C_{t,k}\left(\dfrac{\log(2(d+1))}{n}\right)^{t/(k+5)}, & \text{if } t<s,
    \end{cases}
  \]
  where $\bar C_{t,k}$ depends on $t$, $k$, $G$, $B$, $\CX$, $\CK$, $\|p\|_\infty$, 
 and the constants $C_s$ and $C_k$ from Lemma~\ref{lem:matern}, 
  but is independent of the ambient dimension $d$.
\end{proposition}
If $g$ is Lipschitz continuous, then
\(
  \mathbb E\left[\mathcal R_\ell(\hat f_D^{(1)})-\mathcal R_\ell^*\right]
  =
  O\left(\left(\frac{\log(2(d+1))}{n}\right)^{1/(k+5)}\right).
\)
As in the latent-subspace case in Proposition~\ref{prop:adaptivity}, the rate depends on an intrinsic dimension rather than directly on the ambient dimension. The assumptions are different, however: the earlier result allows an arbitrary low-dimensional linear projection, whereas the present $\ell_1$ geometry targets coordinate sparsity through the active set $J$.

\subsection{Convergence Rates for NNOWL-Kernel}

The preceding results concern fully trained neural networks, whose rate analysis uses the variation complexity $V_1$ in \eqref{eqn:variation_norm}. We now turn to the kernel formulation introduced in Section~\ref{sec:kernel}. Because its hidden-layer distribution is fixed, its natural complexity measure is the Hilbertian norm $V_2$ in \eqref{eqn:hilbertian_norm}. 
For the rate analysis, we use the constrained counterpart of the penalized NNOWL-Kernel estimator in \eqref{eq:kernel-owl-functional}. 
The following lemma, based on \citet[Sec.~9.5.3]{bach2024learning}, links $V_2$ to Sobolev smoothness.

\begin{lemma}
  \label{lem:gamma2_upper}
  Under regularity and domination assumptions on the reference measure $\rho_0$, we have \(
    V_2(f)\le C_2\|f\|_{H^s},
  \)
  where $s=d/2+5/2$, and $C_2>0$ depends on the dictionary radius $\CK$, the dimension $d$, and the regularity constants of $\rho_0$.
\end{lemma}

Combining this embedding with the RKHS estimation bound and the Sobolev approximation argument yields the following rates.

\begin{proposition}
  \label{prop:kernel_risk_rate}
  Assume that $\ell$ is $G$-Lipschitz continuous, $R/\pi(A,\bmX)\le B$ almost surely, $f_*\in H^t(\mathbb R^d)$, and assumptions in Lemma~\ref{lem:gamma2_upper} hold. Let $\hat f_{V_2,D}$ denote the empirical risk minimizer over the RKHS ball $\{f:V_2(f)\le D\}$. Set $s=d/2+5/2$. Then, for a suitable choice of $D$, there exist constants $\tilde C_2>0$ and $\bar C_{2,t,d}>0$ such that
  \begin{equation*}
    \mathbb E\left[\mathcal R_\ell(\hat f_{V_2,D})-\mathcal R_\ell^*\right]
    \le
    \begin{cases}
      8\,G\tilde C_2 B\,\|f_*\|_{H^s}n^{-1/2}, & \text{if } t\ge s,\\[1.2ex]
      \bar C_{2,t,d}\,n^{-t/(d+5)}, & \text{if } t<s.
    \end{cases}
  \end{equation*}
  The constant $\tilde C_2$ depends on the covariate radius $\CX$, the dictionary radius $\CK$, $d$, and the constant $C_2$ in Lemma~\ref{lem:gamma2_upper}, while $\bar C_{2,t,d}$ depends on $t$, $d$, $G$, $B$, $\CX$, $\CK$, $\rho_0$, $\|p\|_\infty$, and the constants $C_s$ and $C_k$ from Lemma~\ref{lem:matern}.
\end{proposition}

Proposition~\ref{prop:kernel_risk_rate} yields the same rates as the variation-norm result in Proposition~\ref{prop:risk_rate}. Both attain the parametric rate $n^{-1/2}$ when $f_*$ is sufficiently smooth and the Sobolev rate $n^{-t/(d+5)}$ when $t<s=d/2+5/2$. The underlying function classes and complexity measures nevertheless differ. Proposition~\ref{prop:risk_rate} controls fully trained networks through the variation norm $V_1$, which bounds the total variation of a signed coefficient measure and permits sparse or singular ridge representations. In contrast, Proposition~\ref{prop:kernel_risk_rate} controls the fixed-kernel estimator through the Hilbertian norm $V_2$, which requires an absolutely continuous representation $d\mu=g\,d\rho_0$ with bounded $\norm{g}_{L^2(\rho_0)}$. Hence, although the exponents coincide under the Sobolev comparison and the domination condition on $\rho_0$, the constants and approximation classes differ. In particular, the $V_2$ bound depends on how well $\rho_0$ dominates the coefficients in the ridge-function construction based on Fourier analysis, as required in Lemma~\ref{lem:gamma2_upper}, and the kernel result does not by itself capture the structural adaptivity of fully trained networks.

\subsection{Global Convergence and Implicit Bias of NNOWL}\label{sec:global}
NNOWL minimizes the surrogate-risk functional
\(
  \mathscr{R}(f)=\mathbb{E}\left[\frac{R}{\pi(A, \bmX)}\ell(Af(\bmX))\right],
\)
where the expectation may denote either empirical or population risk. Write the parameters of neuron $j$ as $\bmtheta_j=(\zeta_j, \bw_j, b_j) \in \mathcal W=\mathbb R^{d+2}$, and define the corresponding feature map
\(
  \Phi(\bmtheta_j):\bx \mapsto \zeta_j\,\sigma(\bw_j^\top \bx + b_j).
\)
For a network of width $m$, the scaled prediction function is
\(
  f_m=\frac{1}{m}\sum_{j=1}^m \Phi(\bmtheta_j).
\)
Denote \(\bmtheta=(\bmtheta_1,\ldots,\bmtheta_m)\), the finite average can be represented through the empirical measure
\(
  f_m=\int_{\mathcal W}\Phi(\bmtheta)\,d\mu_m(\bmtheta),
\)
where \(
  \mu_m=\frac{1}{m}\sum_{j=1}^m\delta_{\bmtheta_j}
\)
and $\delta_{\bmtheta_j}$ is the Dirac measure at $\bmtheta_j$. The corresponding finite-network objective is
\(
  \mathscr R(f_m) = \mathscr R \left(\int_{\mathcal W}\Phi(\bmtheta)\,d\mu_m(\bmtheta)\right).
\)
Replacing the empirical measure $\mu_m$ by a general probability measure $\mu$ on $\mathcal W$ gives the measure-based objective
\(
  F(\mu)=\mathscr R\left(\int_{\mathcal W}\Phi(\bmtheta)\,d\mu(\bmtheta)\right).
\)
When $\ell$ is convex, $F$ is convex as a function of the measure $\mu$. In the infinite-width, continuous-time idealization, gradient descent on the network parameters is described by a Wasserstein gradient flow of $\mu$ that decreases $F$. Convexity in $\mu$, however, does not by itself guarantee global convergence, because the Wasserstein geometry may still admit nonglobal stationary points. The first consequence of the mean-field theory is that, under a sufficiently rich initialization, these nonglobal stationary points are avoided in the limiting flow.

\begin{proposition}
With ReLU activation, suppose the support of the initial parameter distribution includes all directions in $\mathcal W$. If the resulting Wasserstein gradient flow converges weakly to a distribution, then the limiting distribution is a global minimizer of $F$.
\end{proposition}
One way to spread mass in all directions is to use either a uniform initialization on the sphere or a Gaussian initialization. Together with the homogeneity mechanism of the ReLU activation, this initialization condition preserves the global-optimality argument for NNOWL, even though the outcome weighting changes the surrogate risk being minimized.

After establishing this global-convergence picture, the same mean-field viewpoint can also be used to describe the implicit bias of gradient descent. Consider the unpenalized version of \eqref{eq:erm} for NNOWL with logistic or exponential loss, trained by gradient descent for $t$ iterations. In a fully trained network, taking both the width and the training time to infinity leads to a variation-norm maximum-margin limit.
\begin{proposition}
Let $f_{m,t}=m^{-1}\sum_{j=1}^m\Phi(\bmtheta_j(t))$ denote the network prediction at training time $t$. As $m,t\to+\infty$, suppose the 
predictions $f_{m,t}$ 
associated with the gradient flow
converge up to normalization, 
then the limit is a data-separating function with minimum variation norm \(V_1(f)\).
\end{proposition}

The two informal statements above adapt results from \citet{bach2022gradient} to NNOWL. Formal statements, assumptions, proofs based on mean-field theory \citep{chizat2018global, chizat2020implicit} are deferred to the Supplementary Material.

\section{Simulations}\label{sec:sim}
We evaluate the logistic-loss-based neural-network methods developed in Section~\ref{sec:est} using simulated data. We first illustrate the behavior of the NNOWL algorithms and then compare the NNOWL and NNRWL extensions with competing individualized treatment-learning methods from the literature. The evaluation criterion is the value function $V(\Dcal)$, estimated by
\(
  \widehat V(d)=
  \frac{\mathbb{P}_{n}^{*}\left[\mathbb{I}\{A=d(\bmX)\}R/\operatorname{Pr}(A)\right]}
  {\mathbb{P}_{n}^{*}\left[\mathbb{I}\{A=d(\bmX)\}/\operatorname{Pr}(A)\right]},
\)
where $\mathbb{P}_{n}^{*}$ denotes the empirical average over the evaluation data set \citep{murphy2001marginal}. In all simulations, treatment is assigned as $A\in\{-1,1\}$ with $P(A=1)=0.5$, and $a\in\{-1,1\}$ denotes the observed treatment value. 
Hyperparameters for the computational algorithms are summarized in the Supplementary Material.
\subsection{Illustrative Examples}
\noindent We use the following examples to illustrate overparameterization, kernel approximation, and nonlinear variable selection. 

\noindent\textbf{Example 1:}
This example generates a univariate covariate $x$ uniformly on $[-1,1]$. The response is log-normal with
\(
  \log(R) \sim N\bigl(\mu_A(x),1\bigr).
\)
Let
\(
  b_{0,1}(x)=x, \quad
  \Delta_1(x)=4\Bigl|x+1-0.25-\lfloor x+1-0.25\rfloor-\frac12\Bigr|-1.
\)
We define
\(
  \mu_a^{(1)}(x)=b_{0,1}(x)+a\Delta_1(x).
\)
Here $\lfloor\cdot\rfloor$ denotes the floor function. This modified nonsmooth one-dimensional experiment follows \citet{bach2024learning}, with treatment contrast $\mu_1^{(1)}(x)-\mu_{-1}^{(1)}(x)$ serving as both the true contrast and the population target under logistic OWL.

\begin{figure}[t]
  \centering
  \includegraphics[width=\textwidth]{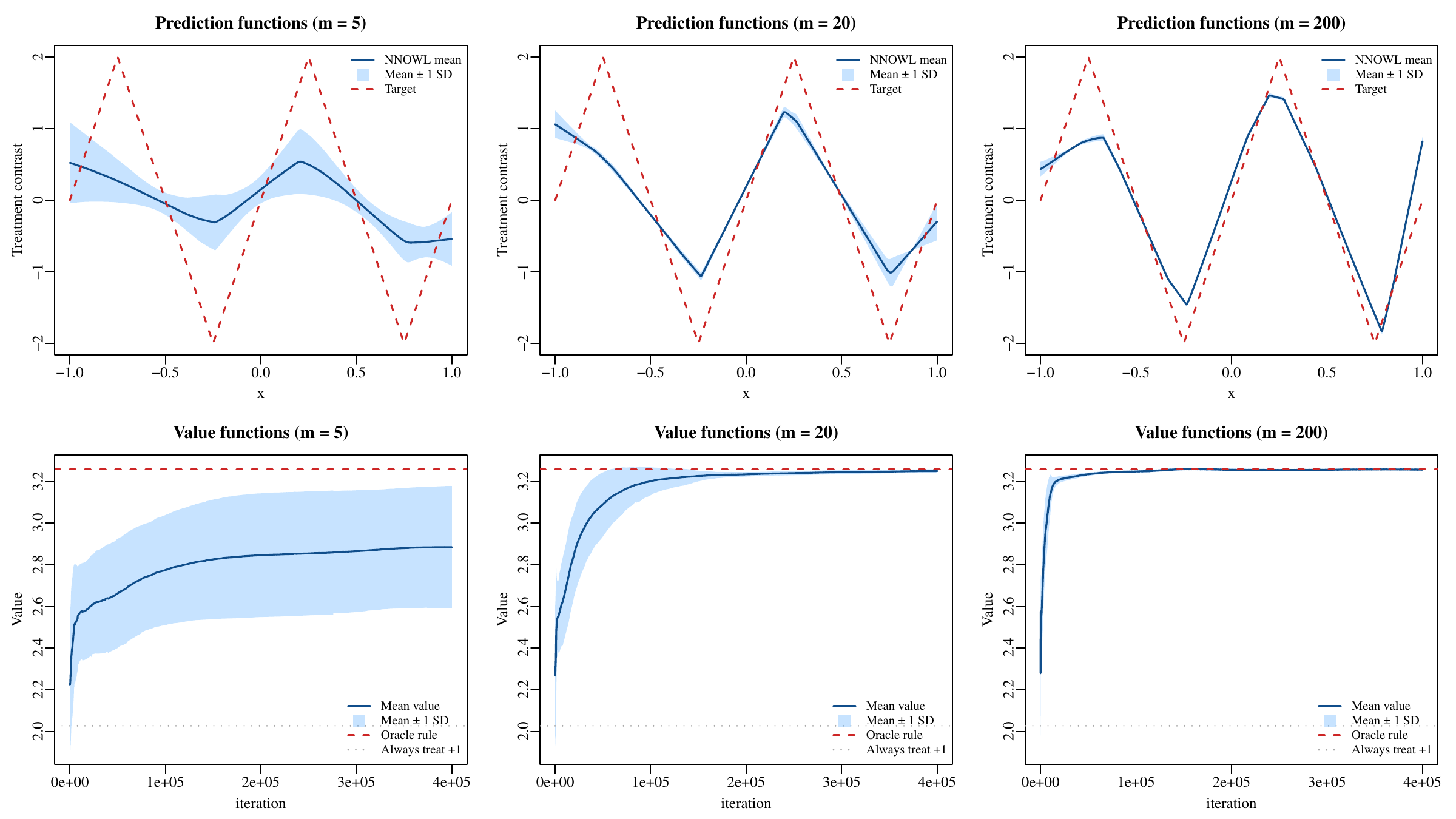}
  \caption{One-dimensional NNOWL illustration. Top: treatment contrast and fitted NNOWL score curves for widths $m \in \{5,20,200\}$. Bottom: mean value functions over replications with a $\pm 1$ standard-deviation band, together with the oracle-rule benchmark and the constant treatment-$+1$ rule.\label{fig:nnowl-1d}}
\end{figure}

We first study the fully trained neural network via NNOWL-Ridge. Specifically, we fit a ReLU NNOWL model by mini-batch stochastic gradient descent to the logistic OWL objective from Section~\ref{sec:nnowl}. For each width $m \in \{5,20,200\}$, we use a training sample of size $n=1{,}000$, an evaluation grid of size $n_{\mathrm{test}}=10{,}000$, and we repeat the experiment over $100$ random replications. The top row of Figure~\ref{fig:nnowl-1d} shows the true treatment contrast together with the fitted NNOWL score curves, whereas the bottom row summarizes the corresponding value functions.
The plot shows the replication mean with a $\pm 1$ standard-deviation band, along with the oracle-rule value and the value of the constant treatment-$+1$ rule. To reduce overfitting, SGD is run in a single-pass regime so that each observation is used only once, with a fresh random initialization in every replication.

Figure~\ref{fig:nnowl-1d} shows a clear benefit of overparameterization. When $m=5$, some fitted score curves fail to track the treatment contrast well, and the resulting value functions fall noticeably below the oracle benchmark. Increasing the width to $m=20$ improves both contrast recovery and decision quality. When $m=200$, the fitted curves are more accurate and stable, the achieved values are closer to optimal, and fewer iterations are needed to reach strong performance. Overall, wider networks yield better and more reliable NNOWL fits in this example.

\begin{figure}[t]
  \centering
  \includegraphics[width=\textwidth]{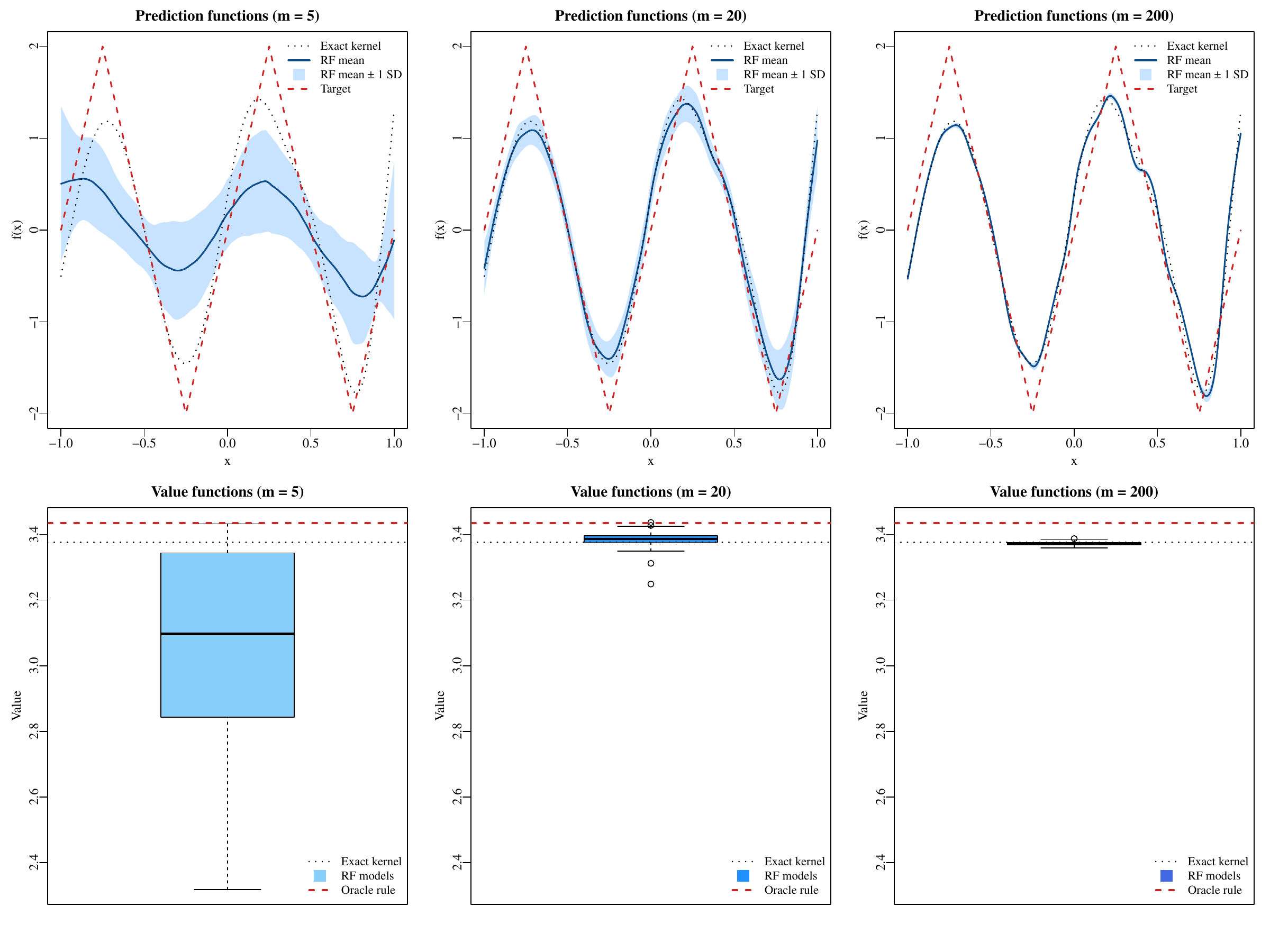}
  \caption{One-dimensional random-feature kernel OWL illustration. Treatment contrast and fitted score curves for widths $m \in \{5,20,200\}$.}
  \label{fig:nnowl-kernel}
\end{figure}

For the kernel function based estimator NNOWL-Kernel, the input weights are generated by first sampling $\tilde{\bw}_j\sim N(0,I_d)$ and then normalizing to $\bw_j=\tilde{\bw}_j/\|\tilde{\bw}_j\|_2$, so that $\bw_j$ is uniformly distributed on the unit sphere. In the one-dimensional setting, this reduces to $\bw_j\in\{-1,1\}$ with equal probability, and the biases are sampled independently as $b_j\sim \mathrm{Unif}[-1,1]$.

The simulation results in Figure~\ref{fig:nnowl-kernel} show how the empirical kernel OWL estimator approaches the exact kernel function as the number of random features increases. For small widths such as $m=5$, the fitted curves exhibit substantial variability across random feature draws and can deviate noticeably from both the exact kernel estimator and the target contrast. As $m$ increases to $20$ and then $200$, the mean random-feature estimator moves progressively closer to the exact kernel fit, while the variability across repetitions decreases. This pattern is consistent with the almost sure convergence of the empirical kernel $\hat{\mathsf{k}}_m(x,x')$ to the population kernel $\mathsf{k}(x,x')$ as $m\to\infty$.

The boxplots of the achieved values provide a complementary summary of decision performance across repeated simulations. When $m=5$, the values are more dispersed and typically lie farther below the benchmark set by the exact kernel estimator. As the number of random features increases, the boxplots become tighter and shift upward, indicating that the fitted treatment rules become both more stable and more effective. By $m=200$, the value distribution is much closer to that of the exact kernel method, with substantially reduced Monte Carlo variability. These boxplots therefore reinforce the conclusion from the fitted curves: increasing the number of random features improves both the accuracy of the estimated contrast function and the quality of the resulting treatment rule. With sufficiently many random features, the empirical kernel estimator provides a computationally efficient approximation to the infinite-width RKHS solution while retaining strong predictive and decision performance.

In this example, for both full neural-network learning and kernel function based estimation, the gains from overparameterization in value functions diminish beyond a certain width. Even so, overparameterization yields more accurate and interpretable estimates of the treatment contrast and, more broadly, of the decision function. Figure~\ref{fig:nnowl-kernel} shows that the kernel function based estimator is more smooth than the full neural network estimators in \ref{fig:nnowl-1d} since the former belongs to a Hilbert space while the latter is in a Banach space $\mathcal{F}_1$.

\begin{figure}[t]
  \centering
  \includegraphics[width=\textwidth]{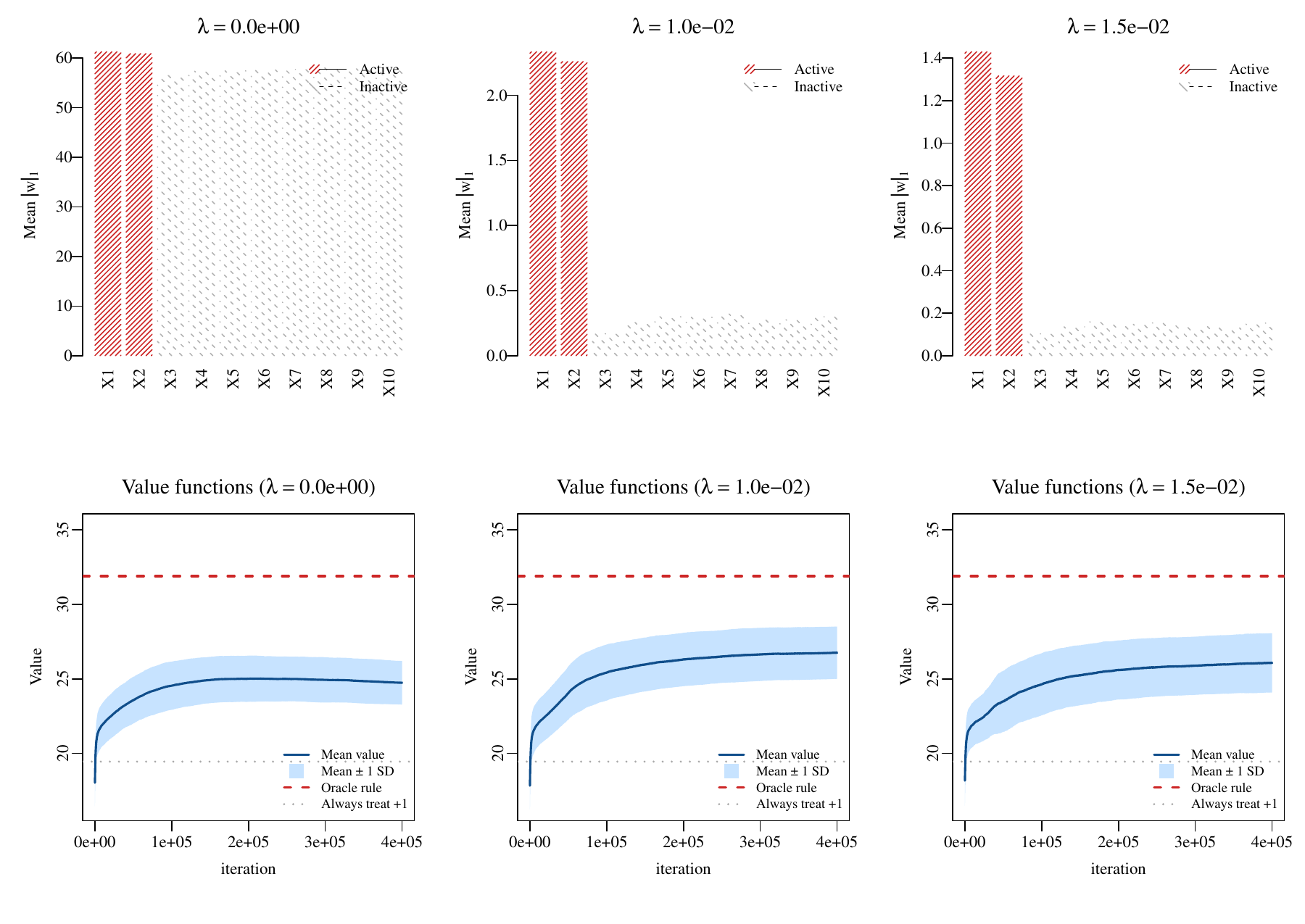}
  \caption{
  Nonlinear variable selection of NNOWL. Top: mean value of $\|\bw\|_1$. Bottom: mean value functions over replications with a $\pm 1$ standard-deviation band, together with the oracle-rule benchmark and the constant treatment-$+1$ rule.\label{fig:nnowl-vs}}
\end{figure}

\noindent\textbf{Example 2:}
We generate covariates $x=(x_1,\ldots,x_d)$ with $d=10$, where each coordinate is sampled independently from $[-1,1]$.
As in Example~1, the response is log-normal,
\(
  \log(R)\sim N\bigl(\mu_A^{(2)}(x),1\bigr).
\)
Let
\(
  b_{0,2}(x)=1+x_1+x_2+2x_3+0.5x_4,\quad
  \Delta_2(x)=\sign\{0.8-x_1^2-x_2^2\}.
\)
We define
\(
  \mu_a^{(2)}(x)=b_{0,2}(x)+a\Delta_2(x).
\)
Thus only the first two covariates determine the treatment contrast, while the remaining coordinates are irrelevant for the optimal treatment rule. We train the sparse NNOWL-VS logistic model with sample size $n=1,00
0$.
Results are averaged over $100$ random replications and evaluated on an independent test sample of size $n_{\mathrm{test}}=10{,}000$. Figure~\ref{fig:nnowl-vs} shows the impact of regularization parameter $\lambda$.

Figure~\ref{fig:nnowl-vs} illustrates the nonlinear variable-selection behavior of the sparsity-regularized NNOWL estimator. The top panel tracks the average input-weight norm during training and shows how the penalty controls the size of the first-layer weights. This is consistent with the intended role of input-layer regularization: it discourages unnecessary dependence on all coordinates and encourages the fitted network to concentrate its representation on covariates that are useful for treatment assignment. The bottom panel shows that the estimated value function improves with the $\ell_1$ penalty while $\lambda=0.01$ is better than $\lambda = 0.015$.
Overall, the figure suggests that input-layer $\ell_1$ regularization can provide meaningful nonlinear variable selection while improving the decision making.

\subsection{Comparative Analysis}

We employ the simulation settings described in \citet{jiang2024deep}, using synthetic randomized-treatment data across low- and high-dimensional scenarios. For each example, $X_1,\ldots,X_p$ are generated independently from $\mathrm{Unif}(-1,1)$ and
\(
  R\mid X=x,A=a \sim N\bigl(\mu_a(x),1\bigr),
\)
where the treatment-specific means are defined by the baseline and contrast functions summarized in Table~\ref{tab:simulation_settings}.

\begin{table}[htbp]
\centering
\caption{Simulation examples.}
\label{tab:simulation_settings}
\footnotesize
\begin{threeparttable}
\setlength{\tabcolsep}{3pt}
\begin{tabular}{@{}ccp{3.6in}p{1.75in}@{}}
\toprule
Examples & $d$ & $b_0(x)$ & $\Delta(x)$ \\
\midrule
3--4 & 5 & $0.5+0.5x_1+0.8x_2+0.3x_3-0.5x_4+0.7x_5$ & $0.2-0.6x_1-0.8x_2$ \\
5--6 & 25 & $0.5+0.6x_1+0.8x_2+0.3x_3-0.5x_4+0.7x_5$ & $0.6-x_1^2-x_2^2$ \\
7--8 & 100 & $1+0.6x_1+0.8x_2+0.3x_3-0.5x_4+0.7x_5$ & $\begin{aligned}[t]{}&0.45-0.1x_1^2-0.2x_2^2+0.3x_3^2\\[-0.2em]&{}+0.2x_4^2-0.9x_5^2\end{aligned}$ \\
9--10 & 800 & $\begin{aligned}[t]{}&0.5+0.6\sum_{j=1}^{10}x_j+0.8\sum_{j=11}^{20}x_j+0.3\sum_{j=21}^{30}x_j\\[-0.2em]&{}-0.5\sum_{j=31}^{40}x_j+0.7\sum_{j=41}^{50}x_j+0.5\sum_{j=51}^{60}x_j\\[-0.2em]&{}+0.4\sum_{j=61}^{70}x_j-0.4\sum_{j=71}^{80}x_j+0.2x_{81}-0.9x_{82}\end{aligned}$ & $\begin{aligned}[t]{}&0.6-0.1\sum_{j=1}^{15}x_j^2\\[-0.2em]&{}-0.2\sum_{j=16}^{30}x_j^2\\[-0.2em]&{}+0.3\sum_{j=31}^{45}x_j^2\end{aligned}$ \\
\bottomrule
\end{tabular}
\parbox{0.94\textwidth}{Note: Within each pair, the first example uses the identity link, $\mu_a(x)=b_0(x)+a\Delta(x)$, and the second uses the exponential link, $\mu_a(x)=\exp\{b_0(x)+a\Delta(x)\}$, except Example 10, where $\mu_a^{(10)}(x)=\exp\{b_{0,9}(x)+a\Delta_9(x)-4\}$.}
\end{threeparttable}
\end{table}

The sample size is $n=800$, with $720$ observations used for training and $80$ for tuning; performance is evaluated on an independent test set of size $100{,}000$. We consider dimensions $d\in\{5,25,100,800\}$.
For each framework, we compare three implementations described in Section~\ref{sec:est}: the random-feature-based kernel method (NNOWL-Kernel and NNRWL-Kernel), the $\ell_2$-penalized ridge method (NNOWL-Ridge and NNRWL-Ridge), and the $\ell_1$-penalized variable-selection method (NNOWL-VS and NNRWL-VS).
For RWL, residuals are obtained from main-effect estimates fitted by weighted least squares when $d=5$ and by ridge regression otherwise, with the ridge tuning parameter selected by generalized cross-validation \citep{zhou2017residual,hastie2009elements}. For NNOWL, when the training rewards include negative values, we enforce nonnegative weights by subtracting the minimum training reward from all training rewards. For $d=25$ in Examples 5 and 6, we use a quadratic basis expansion as a feature-engineering step, augmenting the original covariates with their squared terms and increasing the predictor dimension to $50$. 
The final regularization parameter $\lambda$ is selected by maximizing the estimated value on the independent tuning set. For $d=800$, however, because the estimated value functions can be highly variable in high dimensions with a small tuning set, we instead choose the value of $\lambda$ that minimizes the penalized empirical objective on the tuning set.
This training--tuning split preserves the total sample size $n=800$, making the design comparable to the 10-fold cross-validation setting in \citet{jiang2024deep} while reducing computational cost.

The simulations are repeated 100 times. We summarize the performance in Tables~\ref{tab:lowdim_results} and \ref{tab:highdim_results} using the mean and standard deviation of the estimated value functions. Computing times are summarized in the Supplementary Material. To place the results in context, we use the same examples as \citet{jiang2024deep}: our Examples 3--6 correspond to their low-dimensional Scenarios 1--4, and our Examples 7--10 correspond to their high-dimensional Scenarios 5--8.
Their comparisons include deep-learning policy estimators, regression-based methods, tree- or forest-based methods, kernel-based RWL methods, and cross-fitted augmented inverse probability weighted learning (CAIPWL).
Hence, the comparison below is threefold: first, against the best existing methods reported in \citet{jiang2024deep}; second, against their deep-learning methods specifically; and third, between the proposed NNOWL and NNRWL variants.

In the low-dimensional settings, where the main-effect model can be estimated reliably, residual weighting is beneficial as intended: NNRWL generally improves over its NNOWL counterpart for $d=5$ and $d=25$, with particularly visible gains in the nonlinear Examples 5 and 6. For $d=5$, the best NNRWL values are close to the strongest methods in \citet{jiang2024deep}: NNRWL-Ridge attains $0.99$ and $3.63$ in Examples 3 and 4, compared with the best reported Jiang et al. values $1.00$ and $3.64$, and attains $0.80$ and $3.21$ in Examples 5 and 6, compared with their best reported values $0.82$ and $3.22$. For $d=25$, NNRWL is less competitive for the linear Examples 3 and 4 than the best sparse linear methods in \citet{jiang2024deep}, but improves on their best reported values in the nonlinear Examples 5 and 6: NNRWL-Ridge attains $0.75$ and $3.18$, compared with $0.72$ and $3.12$ reported for the best competing methods. These results also show that the proposed neural-network OWL and RWL variants are competitive with the deep-learning methods in \citet{jiang2024deep}, especially in the nonlinear scenarios where flexible decision boundaries are beneficial.

In the high-dimensional settings, the advantage of residual weighting becomes less pronounced because accurate main-effect estimation is more challenging. In these cases, NNOWL is often comparable to or better than NNRWL. Relative to \citet{jiang2024deep}, our methods are below the best CAIPWL values in the simpler high-dimensional Examples 7 and 8 with $d=100$, but are comparable in Example 9 and improve substantially in Example 10, where NNOWL-VS attains $3.45$ compared with the best reported Jiang et al. value $3.19$. For $d=800$, NNOWL matches or improves on the best reported Jiang et al. values in Examples 7, 9, and 10, with the largest gain in Example 10: NNOWL-Kernel attains $3.45$ compared with $2.48$ for the best method in \citet{jiang2024deep}. The main exception is Example 8, where Jiang et al.'s CAIPWL value $4.93$ for $d=100$ and $4.91$ for $d=800$ remain higher than our best value $4.66$; nevertheless, our best value is still higher than the deep-learning methods reported there.

The standard deviations show a similar pattern. In the low-dimensional examples, the ridge versions of NNOWL and NNRWL are usually among the most stable non-oracle methods, and NNRWL-Ridge combines high mean values with small SDs, especially for $d=5$ and $d=25$. In the high-dimensional examples, SDs remain small for Examples 7--9 across most methods, but become much larger in Example 10, indicating a more difficult learning problem. In that setting, NNOWL tends to be more stable than NNRWL when $d=800$, with NNRWL showing larger SDs as well as lower mean values.

Table~S.3 in the Supplementary Material shows that the kernel version is consistently the fastest, while the ridge and variable-selection neural-network fits can become substantially more expensive when the fitted network or feature representation is enlarged. Because the training size is fixed across these experiments, the observed computing time is driven by several interacting factors, including the covariate dimension, network width, tuning grid and number of iterations when applicable (this needs explanations that kernel methods do not involve gradient decent iterations). See configurations in the Supplementary Material. For instance, the $d=25$ runs for Examples 5 and 6 are relatively slow because these cases use a quadratic feature expansion, increasing the effective input dimension to $50$, and are fit with a wider network ($m=1000$). As a result, these settings require more computation than some nominally higher-dimensional cases, including the $d=100$ settings and several $d=800$ kernel fits that use smaller widths or simpler optimization paths.

Overall, these comparisons suggest that RWL can improve efficiency in low-dimensional regimes where the nuisance main-effect model is reliable, whereas OWL may be preferable in high-dimensional regimes where estimating that nuisance component is difficult. Compared with the full set of methods in \citet{jiang2024deep}, the proposed methods are especially competitive for nonlinear examples and for the most challenging high-dimensional setting. Compared specifically with the deep-learning methods in that paper, the proposed NNOWL and NNRWL variants provide comparable or improved performance in several scenarios, while classical sparse or tree-based methods can remain advantageous for simpler linear or sparse structures.
\begin{table}[!htbp]
\centering
\caption{Low-dimensional simulation results for Examples 3--6, reported as value mean (SD) across replications. Boldface indicates the best non-oracle value.}
\label{tab:lowdim_results}
\begin{tabular}{llcccc}
\toprule
$d=5$ & Method & Example 3 & Example 4 & Example 5 & Example 6 \\
\midrule
 & NNOWL-Kernel & 0.85 (0.03) & 3.57 (0.13) & 0.54 (0.02) & 3.07 (0.14) \\
 & NNOWL-Ridge & 0.89 (0.02) & \textbf{3.63 (0.02)} & 0.55 (0.02) & 3.14 (0.06) \\
 & NNOWL-VS & 0.91 (0.02) & 3.60 (0.13) & 0.54 (0.02) & 3.13 (0.07) \\
 & NNRWL-Kernel & 0.98 (0.03) & 3.61 (0.03) & 0.77 (0.06) & 3.21 (0.10) \\
 & NNRWL-Ridge & \textbf{0.99 (0.01)} & \textbf{3.63 (0.02)} & \textbf{0.80 (0.02)} & \textbf{3.21 (0.03)} \\
 & NNRWL-VS & 0.99 (0.02) & 3.59 (0.13) & 0.79 (0.02) & 3.19 (0.07) \\
 & Oracle & 1.00 (0.00) & 3.66 (0.02) & 0.85 (0.01) & 3.31 (0.02) \\
\midrule
$d=25$ & Method & Example 3 & Example 4 & Example 5 & Example 6 \\
\midrule
 & NNOWL-Kernel & 0.82 (0.03) & 3.46 (0.17) & 0.59 (0.03) & 2.92 (0.14) \\
 & NNOWL-Ridge & 0.87 (0.02) & 3.53 (0.06) & 0.59 (0.03) & 3.00 (0.07) \\
 & NNOWL-VS & 0.87 (0.02) & 3.52 (0.09) & 0.59 (0.04) & 2.96 (0.12) \\
 & NNRWL-Kernel & 0.93 (0.05) & 3.51 (0.09) & 0.72 (0.05) & 3.13 (0.09) \\
 & NNRWL-Ridge & \textbf{0.96 (0.01)} & \textbf{3.58 (0.03)} & \textbf{0.75 (0.02)} & \textbf{3.18 (0.03)} \\
 & NNRWL-VS & \textbf{0.96 (0.01)} & 3.57 (0.06) & 0.73 (0.06) & 3.16 (0.07) \\
 & Oracle & 1.00 (0.01) & 3.66 (0.02) & 0.85 (0.00) & 3.31 (0.02) \\
\bottomrule
\end{tabular}
\end{table}

\begin{table}[!htbp]
\centering
\caption{High-dimensional simulation results for Examples 7--10, reported as value mean (SD) across replications. 
Boldface indicates the best non-oracle value in Example 10.}
\label{tab:highdim_results}
\begin{tabular}{llcccc}
\toprule
$d=100$ & Method & Example 7 & Example 8 & Example 9 & Example 10 \\
\midrule
 & NNOWL-Kernel & 1.21 (0.01) & 4.64 (0.04) & 1.10 (0.02) & 3.12 (0.65) \\
 & NNOWL-Ridge & 1.19 (0.03) & 4.58 (0.12) & 1.08 (0.02) & 3.16 (0.56) \\
 & NNOWL-VS & 1.22 (0.01) & 4.66 (0.02) & 1.10 (0.02) & \textbf{3.45 (0.63)} \\
 & NNRWL-Kernel & 1.22 (0.01) & 4.66 (0.02) & 1.10 (0.02) & 3.15 (0.63) \\
 & NNRWL-Ridge & 1.20 (0.03) & 4.64 (0.03) & 1.10 (0.02) & 3.13 (0.55) \\
 & NNRWL-VS & 1.22 (0.01) & 4.66 (0.02) & 1.10 (0.02) & 3.30 (0.71) \\
 & Oracle & 1.32 (0.01) & 5.08 (0.02) & 1.13 (0.02) & 3.67 (0.48) \\
\midrule
$d=800$ & Method & Example 7 & Example 8 & Example 9 & Example 10 \\
\midrule
 & NNOWL-Kernel & 1.22 (0.01) & 4.58 (0.07) & 1.10 (0.01) & \textbf{3.45 (0.58)} \\
 & NNOWL-Ridge & 1.22 (0.01) & 4.66 (0.02) & 1.09 (0.02) & 3.22 (0.81) \\
 & NNOWL-VS & 1.22 (0.01) & 4.66 (0.02) & 1.10 (0.01) & 3.42 (0.73) \\
 & NNRWL-Kernel & 1.21 (0.04) & 4.66 (0.02) & 1.10 (0.01) & 2.97 (0.95) \\
 & NNRWL-Ridge & 1.20 (0.05) & 4.65 (0.05) & 1.10 (0.01) & 2.97 (0.93) \\
 & NNRWL-VS & 1.21 (0.04) & 4.66 (0.02) & 1.10 (0.01) & 3.06 (0.99) \\
 & Oracle & 1.32 (0.01) & 5.08 (0.02) & 1.13 (0.01) & 3.65 (0.55) \\
\bottomrule
\end{tabular}
\end{table}

\section{Data Application}\label{sec:data}

We use data from a randomized trial to develop data-driven individualized treatment rules for asymptomatic individuals at risk for Alzheimer's disease \citep{sperling2023trial}. Because treatment effects in preclinical Alzheimer's disease may be heterogeneous and difficult to detect from marginal comparisons alone, individualized treatment-rule methods provide a principled framework for learning patient-specific recommendations that target expected clinical benefit while accounting for observed covariates.

We evaluate changes from baseline to week 240 in four clinical outcomes: the Preclinical Alzheimer Cognitive Composite (PACC, $n=824$), Cognitive Function Index Participant and Partner (CFI, $n=819$), Alzheimer's Disease Cooperative Study Activities of Daily Living Prevention Questionnaire Partner (ADL Partner, $n=822$), and Clinical Dementia Rating--Sum of Boxes (CDR-SB, $n=826$). Treatment selection is formulated using Q-learning, outcome-weighted learning, and residual-weighted learning frameworks. Baseline covariates include demographic and clinical characteristics, ApoE4 genotype (apolipoprotein E allele 4), quantitative amyloid positron emission PET measures, and volumetric MRI measures, yielding $d=66$ predictors, including three dummy variables for ApoE4 genotype. Except for PACC, smaller changes indicate better outcomes \citep{sperling2023trial}; these endpoints are therefore multiplied by $-1$ before model fitting, while the estimated value functions are reported on the original scale.

For each endpoint analysis, the sample is divided into training, tuning, and validation sets in a 70/15/15 ratio. Two subjects had missing BMI values, and one subject had missing MRI measures; these values are imputed using medians. Predictor variables are standardized. To avoid data leakage, imputation and standardization are performed using training-set statistics.
In addition to $\ell_1$-penalized Q-learning, we compare kernel-based OWL methods using hinge loss (WSVM) and truncated logistic loss with threshold 1, kernel-based RWL methods using smooth ramp and truncated logistic loss, and neural-network-based OWL and RWL estimators. For RWL, residuals are obtained from ridge regression estimates of the main effect, as in the simulation study. Regularization parameters and kernel or network tuning parameters are selected using the tuning set, and the learned rules are evaluated on held-out validation data using estimated value functions. For single hidden-layer neural networks, we use $m=200$ hidden neurons throughout; we also consider two hidden-layer networks with $m=200$ neurons in the first layer and $m=100$ neurons in the second layer. The same neural-network configurations are used for NNOWL and NNRWL, while the same kernel-method grid-search configurations are applied to the truncated logistic-loss versions of OWL and RWL. Additional implementation details, tuning configurations and results of computing times are provided in the Supplementary Material.


\begin{table}[!htbp]
\centering
\small
\setlength{\tabcolsep}{3pt}
\caption{Estimated value functions for the Alzheimer's disease application across 100 replications, reported as mean (SD).}
\label{tab:a4-results}
\begin{tabular}{@{}lcccc@{}}
\toprule
Method & PACC & ADL Partner & CDR-SB & CFI \\
\midrule
Q-learning & -1.17 (0.45) & -1.74 (0.69) & 0.62 (0.16) & 1.46 (0.58) \\
\midrule
\multicolumn{5}{@{}l}{\textit{Kernel methods}} \\
WSVM-Linear & -1.13 (0.44) & -1.68 (0.69) & 0.62 (0.16) & 1.36 (0.63) \\
WSVM-Gauss & -1.16 (0.47) & -1.83 (0.77) & 0.63 (0.17) & 1.38 (0.59) \\
OWL-BR-Exp & -1.18 (0.38) & -1.83 (0.71) & 0.62 (0.18) & 1.33 (0.63) \\
OWL-BR-Gauss & -1.21 (0.47) & -1.87 (0.71) & 0.60 (0.17) & 1.34 (0.61) \\
OWL-BR-Matern & -1.18 (0.42) & -1.87 (0.77) & 0.60 (0.17) & 1.35 (0.59) \\
RWL-BR-Exp & -1.33 (0.49) & \textbf{-1.98 (0.76)} & 0.63 (0.17) & 1.49 (0.55) \\
RWL-BR-Gauss & -1.30 (0.51) & -1.90 (0.78) & 0.63 (0.19) & 1.46 (0.56) \\
RWL-BR-Matern & -1.30 (0.52) & -1.96 (0.75) & 0.64 (0.19) & 1.49 (0.54) \\
RWL-SR-Linear & -1.17 (0.48) & -1.83 (0.70) & 0.66 (0.17) & 1.56 (0.58) \\
RWL-SR-Gauss & -1.33 (0.51) & -1.89 (0.74) & \textbf{0.60 (0.16)} & 1.45 (0.57) \\
\midrule
\multicolumn{5}{@{}l}{\textit{Neural networks}} \\
NNOWL-Kernel & -1.17 (0.53) & -1.78 (0.67) & 0.63 (0.15) & 1.33 (0.61) \\
NNOWL-Ridge & -1.11 (0.49) & -1.75 (0.70) & \textbf{0.60 (0.16)} & 1.32 (0.67) \\
NNOWL-VS & \textbf{-1.10 (0.48)} & -1.82 (0.71) & \textbf{0.60 (0.16)} & \textbf{1.28 (0.62)} \\
NNRWL-Kernel & -1.18 (0.50) & -1.77 (0.68) & 0.65 (0.17) & 1.53 (0.54) \\
NNRWL-Ridge & -1.12 (0.46) & -1.74 (0.69) & 0.68 (0.19) & 1.55 (0.57) \\
NNRWL-Ridge-2Layer & -1.13 (0.49) & -1.73 (0.68) & 0.68 (0.17) & 1.62 (0.57) \\
NNRWL-VS & -1.16 (0.47) & -1.88 (0.64) & 0.67 (0.17) & 1.48 (0.52) \\
\bottomrule
\end{tabular}

\vspace{0.3em}
\begin{minipage}{0.98\linewidth}
\footnotesize
PACC: Preclinical Alzheimer Cognitive Composite; ADL Partner: Alzheimer's Disease Cooperative Study Activities of Daily Living Prevention Instrument; CDR-SB: Clinical Dementia Rating--Sum of Boxes; CFI: Cognitive Function Index; BR: binomial robust; SR: smooth ramp; Exp: exponential; Gauss: Gaussian. Higher values are preferred for PACC; smaller values are preferred for ADL Partner, CDR-SB, and CFI. Bold indicates the best value.
\end{minipage}
\end{table}

Table~\ref{tab:a4-results} reports the estimated validation-set values for the four week-240 outcomes, with PACC listed first as the primary endpoint. Among the neural-network methods, NNOWL performs more favorably than NNRWL for PACC, CDR-SB, and CFI: NNOWL-VS gives the best PACC and CFI values and ties for the best CDR-SB value, whereas the best NNRWL variants are less competitive on these endpoints. For ADL Partner, however, the residual-weighted methods are stronger; NNRWL-VS improves on the NNOWL variants, although the best overall value is achieved by the kernel-based RWL-BR-Exp method. The reported standard deviations are broadly comparable across methods within each endpoint, indicating that most performance differences are differences in average value rather than large differences in replication-to-replication stability. NNOWL and NNRWL have similar variability for PACC and ADL Partner, while NNRWL tends to show slightly smaller standard deviations for CFI and slightly larger standard deviations for CDR-SB. 

\section{Conclusions}\label{sec:dis}
This article systematically investigated neural-network formulations of outcome weighted learning and residual weighted learning, with both ridge-type regularization and nonlinear variable-selection regularization, as well as kernel-based computation. The theory highlights a key distinction: kernel methods optimize over fixed feature representations, whereas fully trained neural networks learn hidden-layer representations jointly with output weights. We also established global convergence and implicit bias for gradient descent in NNOWL using mean-field theory. Together, these results illustrate a trade-off between computational tractability and statistical expressivity: kernel methods lead to simpler optimization and rate analysis once the kernel is fixed, while fully trained networks solve a harder nonconvex problem but can adapt to low-dimensional nonlinear structure that may not be captured by a prespecified kernel.

The empirical results show that NNOWL and NNRWL are competitive across simulations and the A4 Study analysis. NNOWL can outperform NNRWL when residualization depends on a noisy or misspecified main-effect model, whereas RWL can be advantageous when main-effect adjustment stabilizes the weighted classification problem. Future work includes convergence theory for stochastic optimization and deeper architectures for neural-network OWL and RWL, extensions to multi-arm treatment regimes based on weighted multicategory classification \citep{zhang2020multicategory}, and neural-network methods for multi-stage dynamic treatment regimes, building on dynamic-regime OWL ideas such as \citet{zhao2015new}.

\phantomsection\label{supplementary-material}
 \bigskip
 
 \begin{center}
 
 {\large\bf SUPPLEMENTARY MATERIAL}
 
 \end{center}

The supplement contains proofs, global convergence and implicit bias of NNOWL, technical configurations, and computing-time results. \paragraph*{Data Availability Statement} The A4 Study data are available at \url{https://www.a4studydata.org} after registration and
agreement to the terms of use; R code for the data analysis and the associated R package will be made available on CRAN upon publication. \paragraph*{Disclosure Statement} The author reports there are no competing interests to declare. \paragraph*{Acknowledgment} We acknowledge the A4 and LEARN Study teams, funders, coordinating centers, data-sharing partners, and study participants and partners; full acknowledgments are provided in the Supplementary Material. Google Gemini and OpenAI Prism were used for coding assist, proofread and idea exploration during manuscript preparation.

\begin{spacing}{1} 
\begingroup 
\renewcommand{\newline}{}
\renewcommand{\harvardurl}[1]{}
\bibliography{../wangres}
\endgroup
\end{spacing}

\includepdf[pages=-]{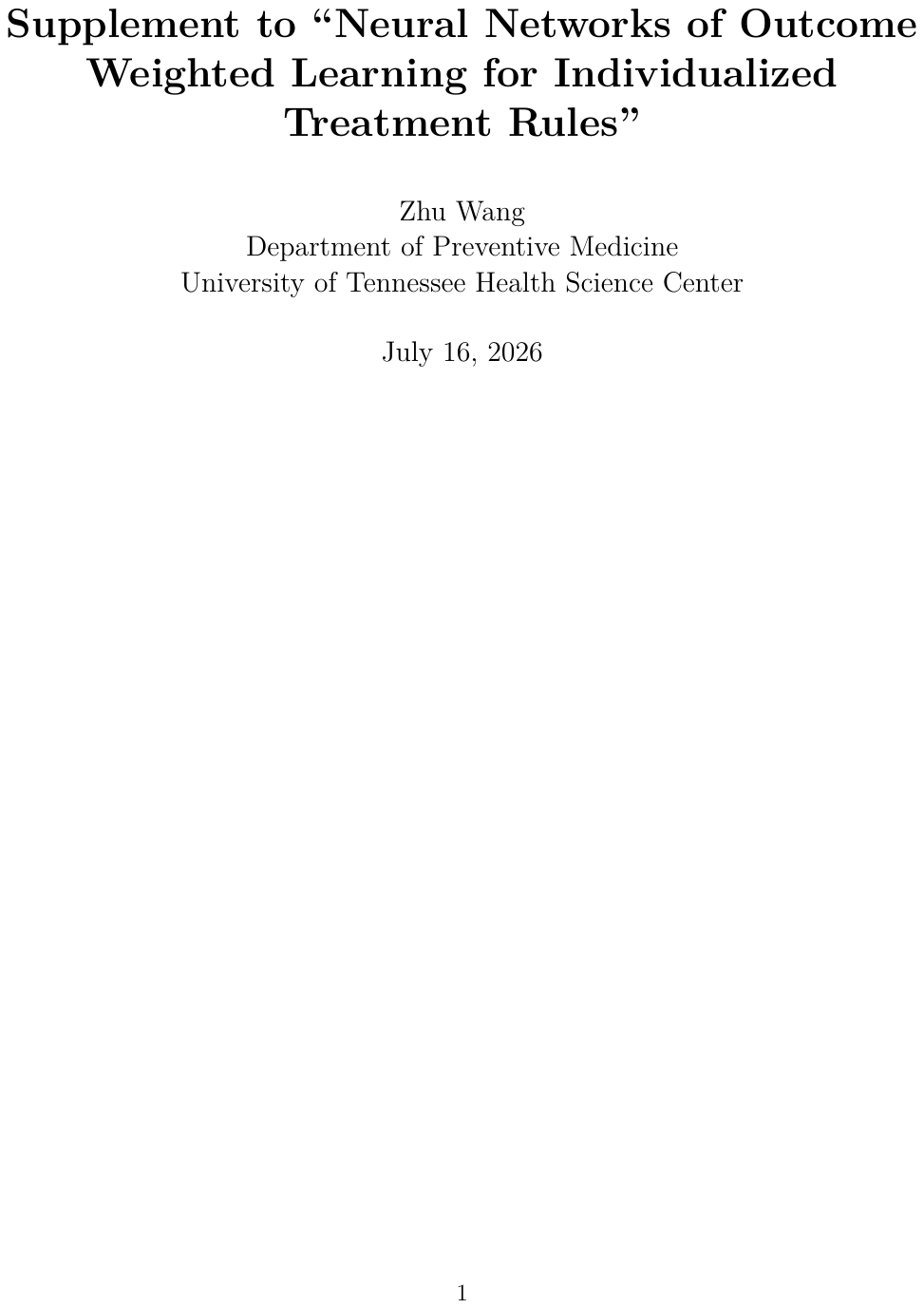}

\end{document}